\documentclass[tightenlines,eqsecnum,floats,aps,amsmath,amssymb,nofootinbib,prd,shownopacs,notitlepage,nobibnotes]{revtex4-1}

\usepackage{amsmath,amssymb,amsfonts,dcolumn,color,graphicx,graphics,latexsym}
\usepackage[mathscr]{eucal}
\usepackage[latin1]{inputenc}
\usepackage{enumerate}
\newcommand{\be}{\begin{equation}}
\newcommand{\ee}{\end{equation}}
\newcommand{\ba}{\begin{eqnarray}}
\newcommand{\ea}{\end{eqnarray}}

\newcommand{\lp}{l_{\mathrm{Pl}}}

\usepackage{epstopdf}

\newcommand{\bmult}{\nopagebreak[3]\begin{multline}}
\newcommand{\emult}{\end{multline}}

\def\rcr{\rho_{_\mathrm{max}}^{^\mathrm{flat}}}

\def\lp{\ell_{\mathrm{Pl}}}
\def\be{\begin{equation}}
\def\ee{\end{equation}}

\begin{document}
\title{Implications of quantum ambiguities in $k=1$ loop quantum cosmology: \\distinct quantum turnarounds and the super-Planckian regime}

\author{John L. Dupuy$^\star$ and Parampreet Singh$^\dagger$}
\affiliation{$^\star$ Department of Physics and Astronomy, University of North Carolina at Chapel Hill, Chapel Hill, North Carolina, USA \\ $^\dagger$ Department of Physics and Astronomy, Louisiana State University,
Baton Rouge, Louisiana, USA \\}

\begin{abstract}
The spatially closed Friedmann-Lema\^{i}tre-Robertson-Walker model in loop quantum cosmology admits two inequivalent consistent quantizations: one based on expressing the field strength in terms of the holonomies over closed loops, and, another using a connection operator and open holonomies. Using the effective dynamics, we investigate the phenomenological 
differences between the two quantizations for the single fluid and the two fluid scenarios with various equations of state, including the phantom matter. We show that a striking difference between the two quantizations is the existence of two distinct quantum turnarounds, either bounces or recollapses, in the connection quantization, in contrast to a single distinct quantum bounce or a recollapse in the holonomy quantization. These results generalize an earlier result on the existence of two distinct quantum bounces  for stiff matter by Corichi and Karami. However, we find that in certain situations two distinct quantum turnarounds can become virtually indistinguishable.  And depending on the initial conditions, a pure quantum cyclic universe can also exist undergoing a quantum bounce and a quantum recollapse. We show that for various equations of states, connection based quantization leads to super-Planckian values of the energy density and the expansion scalar at quantum turnarounds. Interestingly, we find 
that 
very extreme energy densities 
can 
also occur for the holonomy quantization, breaching the maximum allowed density in the spatially flat loop quantized model.  However, the expansion scalar in all these cases  is bounded by a universal value. 

\end{abstract}

\maketitle

\section{Introduction}
Quantization ambiguities in the quantization of classical spacetimes come in different flavors. Some of them, such as the factor ordering ambiguities, result in quantum physics which is generally quite similar for many phenomenological questions. On the other hand, quantization ambiguities resulting from the way the 
classical Hamiltonian is quantized can result in very different physics. In some cases, the resulting physics can be so different that one may be able to select a particular quantization over the others. Thus, eliminating at least some of the quantization ambiguities using conditions for the viable physics. 
An example of restriction of such quantum ambiguities exist in loop quantum cosmology (LQC), a non-perturbative approach based on Ashtekar variables to quantize cosmological spacetimes using the techniques of loop quantum gravity \cite{as-status}. Here for spatially flat cosmological models, only one particular way 
of quantization of Hamiltonian constraint results in a consistent physics \cite{cs-unique,cs-geom}.

Our goal in this manuscript is to understand some of the qualitatively different physical implications of ambiguities in the  quantization of the Hamiltonian constraint in the 
spatially closed Friedmann-Lema\^{i}tre-Robertson-Walker (FLRW) spacetime in LQC. Due to the underlying quantum gravitational effects, the structure of the 
quantum spacetime in this framework is fundamentally different from the classical general relativity (GR), and also other approaches to quantum cosmology such as the Wheeler-DeWitt quantization.  When the spacetime curvature approaches Planckian value, classical differential geometry of GR is replaced by a discrete quantum geometry. Unlike in the classical theory and the Wheeler-DeWitt framework, the Hamiltonian constraint is not a differential equation, but a quantum difference equation with equal spacings in the volume of the universe. At large volumes, i.e. small spacetime curvature for matter satisfying weak energy condition, the quantum difference equation in LQC is well approximated by the Wheeler-DeWitt equation. The departures between the dynamics in LQC and GR vanish. However, in the Planck regime, the differences between LQC and GR become very significant which have an important bearing on the fate of singularities. 
A striking consequence of quantum geometry is the existence of bounce and the quantum resolution classical big bang singularity shown first using numerical simulations \cite{aps1,*aps2,aps3}, and confirmed for arbitrary states in an exactly solvable model in LQC \cite{slqc}. In this model, consistent quantum probabilities can be computed which show that for arbitrary states the probability for singularity to occur is zero \cite{consistent-lqc,*consistent-vertex}. Notably the bounce can also be captured successfully using an effective spacetime description in LQC, confirmed by a large number of numerical simulations (see Ref. \cite{ps12,khanna-review} for a review). 

Results of the bounce in the quantum theory and resolution of big bang singularity have been verified for a variety of initial states and spacetimes \cite{apsv,closed-warsaw,bp-lambda,kv-open,kp-lambda,ap-lambda,madrid-comp,rad,ps12,numlsu-2,*numlsu-3}.
Investigations of various cosmological spacetimes show that LQC results in GR at infra-red scales and successfully resolves the singularity problem of the latter by the quantum gravity modifications at the ultra-violet scales. 
Perhaps the best evidence of a precise agreement of LQC with GR at infra-red scales and the non-trivial role of quantum gravity in singularity resolution at ultra-violet scales  appears in the spatially closed FLRW model. Classically the spacetime recollapses and encounters a big crunch singularity. For such a matter, recollapse occurs unless strong energy condition is violated and there exists sufficiently strong cosmic acceleration at late times. The quantization of the $k=1$ FLRW model in LQC results in a cyclic universe, avoiding big bangs and big crunches at the Planck scale and giving a precise agreement with the classical theory for volume of the universe at which the recollapse occurs \cite{apsv}. Numerical simulations with states which are sharply peaked initially at large volumes show that the fluctuations of the quantum state remain well bounded for a large number of cycles. A result in agreement with the bounds on variation of relative fluctuation of the quantum state in non-singular evolution for 
spatially flat models in LQC \cite{recall,*kp-fluc,*cm-fluc1,*cm-fluc2}.

How is this physics at Planck scale affected by the ambiguities in the quantization procedure, especially the way to obtain the quantum Hamiltonian constraint from its classical counterpart? As remarked earlier there are very strong constraints for the spatially flat models in LQC. These arise if one demands that the singularity resolution occurs at a well defined curvature scale, infra-red limit agrees with GR, and that physics is independent of any fiducial structures used to define the symplectic structure. For the spatially flat isotropic model there is only a unique quantization, the improved dynamics \cite{aps3}, which satisfies this criteria \cite{cs-unique,cs-geom}. The situation becomes subtle in the presence of the  spatial curvature and anisotropies. Let us first recall the status in the anisotropic models. For Bianchi-I models, if the spatial topology is compact then two possible quantizations exist broadly satisfying the above criteria leading to a similar physics except in certain special 
situations \cite{ps16a}. The picture is similar for the loop quantization of anisotropic 
models 
with a non-zero spatial curvature. These include Bianchi-IX \cite{we-bianchi9,pswe,ck-b9} and Kantowksi-Sachs spacetimes \cite{bv,cs-schw}, where again two different consistent quantizations exist but with qualitatively different physics \cite{ck-b9,bgps-spatial,cs-schw,ks-constant,*ks-bound,*ks-strong}. Interestingly, to loop quantize anisotropic models in presence of the spatial curvature, departures from the 
original procedure used in LQC are needed. In the standard procedure, the field strength of the Ashtekar-Barbero connection in the Hamiltonian constraint is expressed in terms of holonomies on a closed loop. However, due to the complexities of interplay of spatial curvature and  anisotropies, defining holonomies in terms of closed loops is very non-trivial. To overcome this obstacle one defines a non-local connection operator using open holonomies which is then used to construct the quantum Hamiltonian constraint \cite{awe-bianchi2}. In our analysis, we will refer to this approach as the connection operator quantization or the connection quantization approach.


The connection operator method leads to a consistent quantization of $k=1$ FLRW model in LQC as is shown by Corichi and Karami \cite{ck-closed1}. This quantization is unique from the earlier consistent quantization of the same spacetime using closed holonomies \cite{apsv,closed-warsaw}. At the coarse level, both the quantizations yield similar physics in the following sense. They are both non-singular and result in classical GR at small spacetime curvatures, including the correct volume at which classical recollapse occurs. But, important differences exist, some of which have been noted earlier. First, it was observed that unlike the holonomy quantization approach, the connection operator based quantization results in two distinct bounces and a classical recollapse, shown explicitly for the case of the massless scalar field \cite{ck-closed1}. When a massless scalar field sourced universe bounces at very large volumes, the two distinct bounce volumes approach each other. The situation becomes similar to the 
case of the holonomy quantization of this spacetime with a massless scalar field which yields one distinct quantum bounce and one classical
recollapse of the universe in each cycle. However, the fate of the singularity resolution in the connection and holonomy quantizations, and 
their contrasting behaviors, for more general fluid contents has remained open.\footnote{An exception is the case of study of exotic 
singularities sourced with fluids with exotic equations of states in the $k=1$ model in LQC quantized with the holonomy approach \cite{psvt}.} 
In the presence of fluids with different equations of states, very interesting physics can arise as is discussed in this manuscript. Quantum 
bounces can be absent and replaced by quantum recollapses depending on the equation of state. The second already noted difference between 
the two quantizations deals with the curvature scale at the bounce(s). It has been argued that in the 
connection approach, the expansion scalar and the energy density are not universally bounded. Whereas, the expansion scalar is predicted to be universally bounded for the holonomy case \cite{bgps-spatial}. Note that the effective Hamiltonian in LQC for the spatially closed model gets quantum gravitational modifications from expressing the field strength or the connection in terms of holonomies, as well as expressing the inverse triad or the volume operators as eigenvalues of quantum operators corresponding to commutators of holonomies with positive powers of triads. The latter inverse volume or inverse triad modifications generally play little role in influencing the singularity resolution and quantum dynamics as compared to the holonomy modifications. Nevertheless they play a role 
in making the behavior of energy density and expansion scalar bounded, albeit not universally, in connection quantization \cite{bgps-spatial,ck-closed2}. The pertinent open question is the following: What is the magnitude of difference in the curvature regimes phenomenologically for different matter content? Can the differences be extremely large depending on the matter content and the initial conditions? 

\begin{figure}[]
\centering
\includegraphics[scale=0.45]{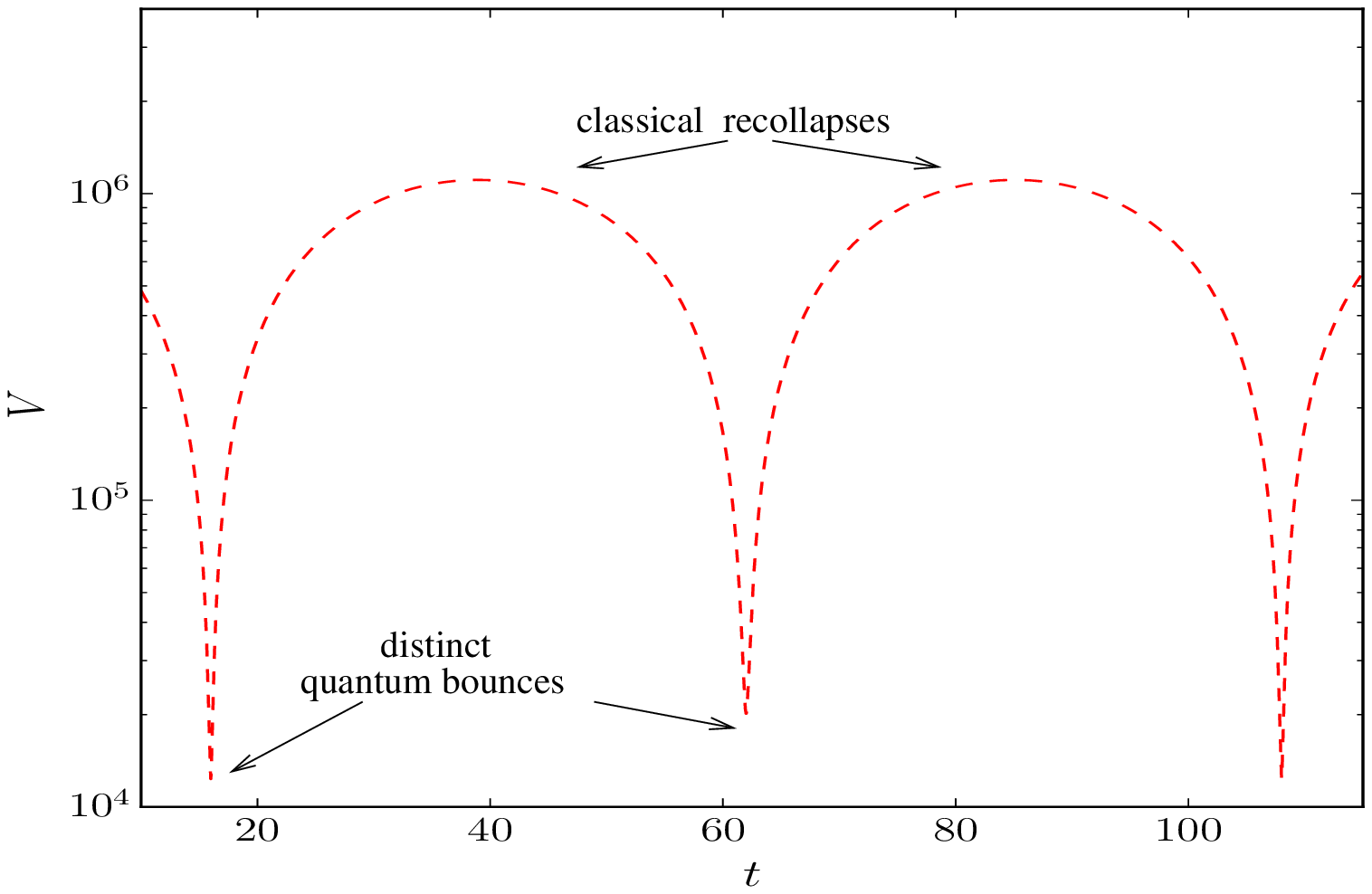}
\includegraphics[scale=0.48]{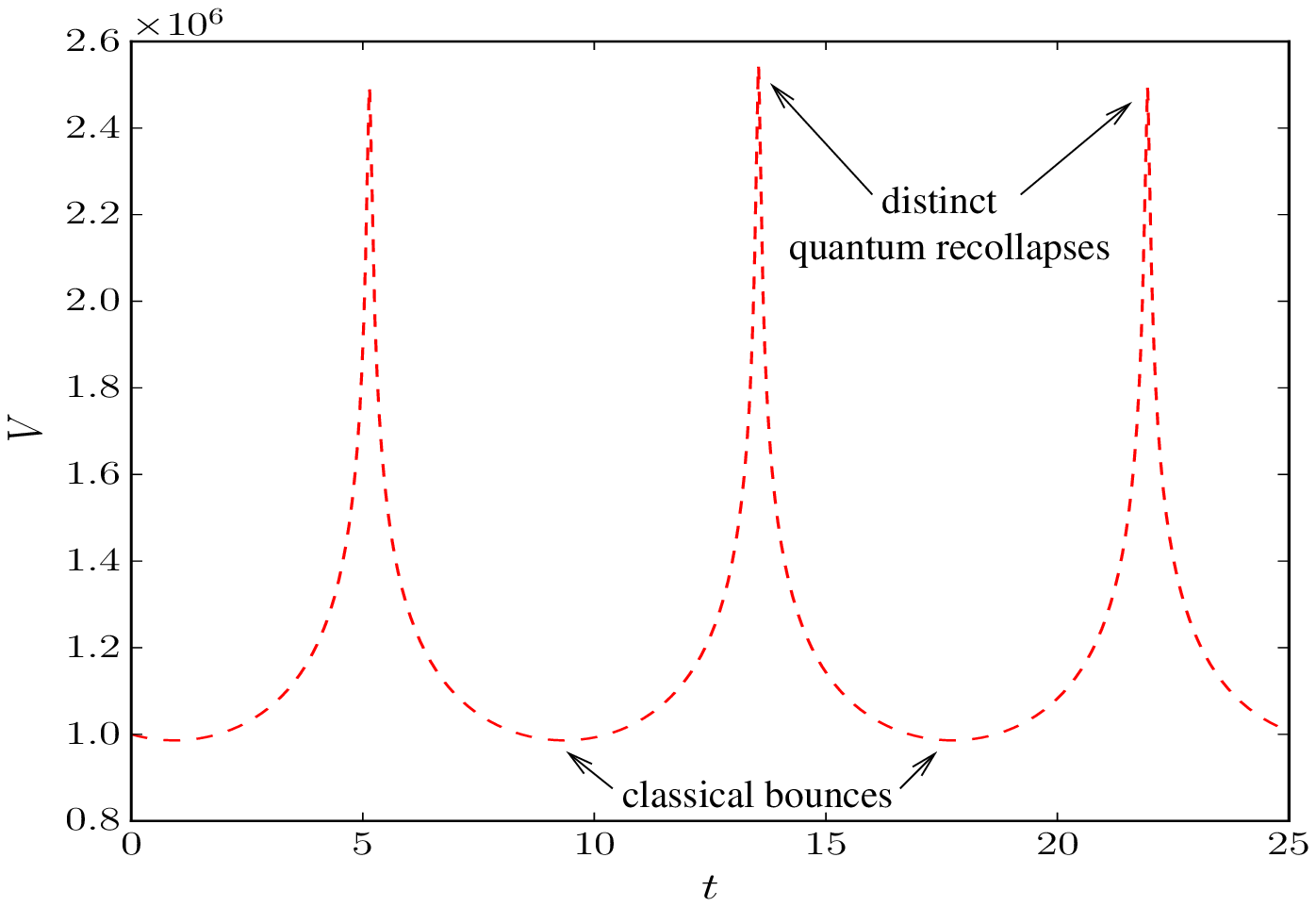}
\caption{A cartoon depiction of the distinct quantum turnarounds for the connection quantization is shown. In the left figure, the evolution is sourced by a matter content obeying the strong energy condition ($w > -1/3$). Such an evolution results in two distinct quantum bounces and a classical recollapse in each cycle. In the right figure, evolution is sourced by a phantom fluid with $w \ll -1$. In this case, quantum turnarounds are no longer bounces but recollapses. The classical turnaround is a bounce in the phantom case. Distinct quantum recollapses become indistinguishable for equations of state closer to negative unity. In the holonomy quantization, there is only a single quantum bounce and a quantum recollapse in the left and the right figures respectively.}
\end{figure}

Answers to above questions are crucial to understand the contrast in the physical implications of the connection and holonomy quantization approaches for the $k=1$ model in LQC. Not only such a study allows gaining understanding of the robustness of the two distinct bounce scenario in the connection quantization, it can also potentially put constraints on the viability of connection quantization approach if the latter results in singularity resolution at high super-Planckian values of curvature. In this paper, we explore answers to above questions by phenomenologically considering matter composed of single and two fluids with a fixed equation of state, allowing equation of state which even violates the weak energy condition. Our numerical investigations are based on assuming the validity of the effective dynamics in the connection and the holonomy quantization approaches.

The main results we obtain on the existence of two distinct quantum turnaround in connection quantization versus one in holonomy quantization for different matter are the following. For the case of single fluid satisfying the strong energy condition (i.e. with equation of state $w > -1/3$), two distinct bounces occur in general in the connection quantization approach (see Fig. 1).  Recall this is the needed value of equation of state in the classical cosmology for decelerated expansion to occur, allowing a classical recollapse in the future. The same result holds for the two fluid scenarios. These results are a generalization of the phenomenological investigation of the massless scalar field case carried out in Ref. \cite{ck-closed1}. The situation for the existence of latter changes when the equation of state is $w < -1/3$. For $-1 \leq w \leq -1/3$, accelerated expansion avoids the classical recollapse in the future, killing the cyclic behavior of $k=1$ model. However, for the phantom fluids which have $w 
< -1$  \cite{ssd}, the 
classical singularity lies in the future at large volumes -- the big rip singularity.  There is no big bang in the past, but a classical bounce. We find that the future singularity  is avoided by the quantum gravitational effects both in the holonomy and the connection quantizations. This result is in synergy with the earlier results on the resolution of the big rip singularity in LQC \cite{sst,psvt,lqc-bigrip3, *lqc-bigrip4}. However, various new features arise.  
For universes which become classical at some point in the evolution and have a phantom equation of state close to negative unity, there is only a single distinct quantum recollapse in each cycle in the connection quantization. Strictly speaking, there are two distinct quantum recollapses but they become indistinguishable. However, for much more negative equation of state, we find two quantum recollapses. (A cartoon depiction of a representative case is shown in Fig. 1). Thus, we find an interesting result that in the connection quantization two distinct quantum turnarounds, quantum bounces for $w > -1/3$, and quantum recollapses for $w \ll -1$, occur.  In contrast, in the holonomy quantization only one distinct quantum bounce/recollapse occurs. The phantom fluid in the quantum realm brings out some more surprises. For universes which remain quantum, there is no classical turnaround but a single distinct quantum bounce and single distinct quantum recollapse. In the connection case, one of the distinct 
turnarounds is a quantum bounce and the other turnaround is a quantum recollapse.  One obtains a cyclic quantum universe due to purely 
quantum geometric effects. The situation is similar in holonomy 
and connection quantization approaches albeit with some important differences on the nature of quantum turnarounds.

Our results on the curvature scales associated with singularity resolution lead to some new insights. Working with the effective Hamiltonian without inverse volume modifications which play little role in dynamics as compared to holonomy modifications, we find that in general the connection operator quantization is associated with bounce densities which are larger than the universal maximum $\rcr$ for the spatially flat model in LQC. Depending on the initial conditions, the bounce density can be extremely large compared to the Planck density. Such bounces occur extremely close to the classical singularity. In this regime, the validity of the effective spacetime description itself can be questioned. On the other hand, the holonomy quantization results in energy density at the bounce in agreement with the universal maximum $\rcr$. Interestingly, a novel result emerges for the first time via numerical investigations. There exist initial conditions for which the bound in energy density for the spatially flat 
model in LQC is breached in the holonomy approach. 
Note that in the absence of inverse volume modifications there is no upper bound on the value of energy density in the holonomy quantization of $k=1$ model \cite{apsv}.  
However, even for such cases the expansion scalar in $k=1$ model in holonomy quantization remains bounded above by its universal value. This result 
suggests that in 
the presence of spatial curvature it 
is only the bound in expansion scalar in holonomy quantization in LQC which makes sense.

As remarked above, in this manuscript we ignore the modifications due to the inverse volume effects. 
 These modifications affect the holonomy and the connection quantization Hamiltonian constraints in the same way. In particular, these modifications 
enter the gravitational part of the constraint through a multiplicative term which is identically unity for volume $V$ greater than $3.39 V_{\rm{Pl}}$, where $V_{\rm{Pl}}$ denotes the Planck volume, and scales as  $V/3.39 V_{\rm{Pl}}$ for smaller volumes \cite{ck-closed2}. The inverse volume  modifications can also affect the matter part of the Hamiltonian constraint, depending on the equation of state of the matter content. If the equation of state is closer to unity (i.e. of the massless scalar field), the dependence of energy density on inverse scale factors is largest and so is the effect of inverse volume modifications. These are not identically unity for larger volumes. However, as in the case of the gravitational part of the constraint, inverse volume terms leave a negligible effect if bounce occurs at volumes much greater than the Planck volume. It is important to note that both the holonomy and connection quantizations are affected in the identical way by the inverse volume modifications if we 
choose 
the same matter content. Hence, a meaningful comparison between the two quantization prescriptions  can be made by ignoring these modifications. As we will discuss in Sec. III, for representative 
simulations the error in excluding these terms is less than $10^{-6} - 10^{-12}$ for the case of equation of state equal to unity, depending on the bounce volume. For lower equations of states, as studied in various simulations, the error is much smaller. Even for the case of extreme Planck densities, such as those discussed in Sec. VC, the error in ignoring these modifications turns out to be small ($\sim 10^{-4}$). It is only in the case when the bounces occur at  volumes less than the Planck volume, that these modifications can play an important role. As we will discuss later in Sec. VB, in the connection quantization in the absence of these modifications bounce volume in some cases can be very small compared to the Planck volume, pushing the energy density and expansion scalar at the bounce to very large values. In the presence of inverse volume modifications, energy density and expansion scalar are bounded in the connection quantization, however the estimated theoretical value assuming validity of 
effective dynamics is super-Planckian \cite{ck-closed2}. The energy density in the holonomy quantization also gets bounded due to inverse volume modifications, 
but the bound is again super-Planckian and breaches the universal bound in the spatially flat case $\rcr$. Our simulations in the deep Planckian regime provide a phenomenological evidence for these such large  values and suggest that though inverse volume effects may alleviate the very extreme values of the energy density and the expansion scalar in the connection quantization, super-Planckian values are reached in the connection  quantization for the energy density and the expansion scalar.  The same conclusion holds for the energy density breaching the $\rcr$ in some cases in the holonomy quantization. Thus, the qualitative results reached for the super-Planckian regime in our manuscript are not affected by ignoring the inverse volume modifications.

This manuscript is organized is as follows. In Sec. II, we start with a summary of the effective Hamiltonians and the resulting dynamical equations from the holonomy and connection operator quantization. These dynamical equations are numerically solved in the subsequent sections where results from our investigations are discussed. For single and two fluid case satisfying the strong energy condition, the existence of two distinct bounces in the connection quantization and a single bounce in holonomy quantization is discussed in Sec. III for various choices of equations of state: stiff matter (massless scalar), radiation and dust. Sec. IV deals with the phantom fluids where the quantum turnaround is a recollapse preventing a future classical big rip singularity. Here we find that for universes which become classical at some point in the evolution in the connection quantization there is only a single quantum recollapse unless the phantom fluid violates weak energy condition significantly. In the case when the 
universe 
remains quantum, instead of the three distinct turnarounds -- two quantum and one classical, there are only two quantum turnarounds. One turnaround is identified with the bounce and another with the recollapse. Results are also contrasted with the holonomy quantization. The two fluid model with a phantom fluid is also discussed. Sec. V deals with  comparing the 
scales at which bounce(s) in the energy density occurs in the connection quantization and the holonomy quantization. A possibility which results in the bounce density in the holonomy quantization greater than 
$\rcr$ is discussed. We conclude with a summary of results in Sec. VI.

\section{Effective Hamiltonian dynamics: Holonomy and connection quantization}
Though the underlying quantum Hamiltonian constraint in LQC is a quantum difference equation, for states which are sharply peaked at late times an effective spacetime description can be derived \cite{vt}. The effective Hamiltonian constraint obtained using the geometric formulation of quantum mechanics, results in a modified effective dynamics which has been tested extremely well for a various spacetimes and matter content using numerical simulations. It captures quantum dynamics very accurately and includes the bounce volume. Notably the effective Hamiltonian of LQC can also be obtained using an inverse procedure. Demanding repulsive nature of gravitational dynamics above a spacetime curvature scale, manifested in form of the quadratic term in the energy density and general covariance uniquely leads to the effective Hamiltonian of LQC \cite{ss-inverse1}.

In the following we summarize the effective Hamiltonian dynamics in the holonomy and the connection quantization approaches. As noted earlier, the effective dynamics gets contribution from two types of terms. One coming from the field strength of the connection, written in terms of the closed or open holonomies, and another from the inverse scale factor terms. The latter result in comparatively much weaker contribution to singularity resolution, and hence are ignored in our analysis. Nevertheless, by themselves they can lead to singularity resolution \cite{st} and an interesting phenomenology (see eg. \cite{thermal}). We also discuss some consequences of including these modifications in the following. For further details of the effective Hamiltonians discussed below, we refer the reader to Refs. \cite{apsv,ck-closed1}. The effective Hamiltonian of the spatially closed model in presence of inverse scale factor effects is discussed in Ref. \cite{ck-closed2}.

\subsection{Holonomy quantization}
Let us first note that due to the presence of the underlying symmetries of the homogeneous and isotropic spacetime, the Ashtekar-Barbero connection $A^i_a$ and the conjugate triad $E^a_i$ are symmetry reduced to $c$ and $p$. These satisfy:
\be
\{c,p\} = \frac{8 \pi G \gamma}{3} ~.
\ee
Here $\gamma$ is the Barbero-Immirzi parameter and has a value $\gamma \approx 0.2375$ fixed by the black hole thermodynamics in LQG \cite{entropy}. 
In the holonomy quantization, the field strength of the connection is expressed in terms of holonomies over the closed loops which are shrunk to the 
minimum eigenvalue of the area operator in LQG. The resulting quantum Hamiltonian constraint is a quantum difference equation with  a uniform spacing in volume. 
The effective Hamiltonian corresponding to the quantum Hamiltonian constraint in the holonomy quantization is \cite{apsv}:
\begin{equation}\label{hcons1}
\mathcal{H}_{\rm{eff}}^{\rm{(hol)}} = -\frac{3}{8 \pi G \gamma^2 \lambda^2} V [\sin^2(\lambda \beta - D) - \sin^2 D + (1+\gamma^2)D^2] + {\cal{H}}_{\mathrm{matt}} ~.
\end{equation}
Here $\lambda^2$ is the minimum allowed area eigenvalue in LQG and has a value of $\lambda^2 = 4(\sqrt{3}\pi \gamma) \lp^2$, where $\lp$ is the Planck length. $D$ is a function defined for convenience as  
\be\label{Deq}
D := (\lambda (2\pi^2)^{1/3})/V^{1/3} . 
\ee
Above the gravitational part of the Hamiltonian is expressed in terms of $\beta$ and $V$ phase space pair which have the following relationship $\beta = c/|p|^{1/2}$ and $V = |p|^{3/2}$. The modulus sign arises due to the two possible orientations of the triad. Without any loss of generality, we choose the positive orientation.

The matter part of the Hamiltonian in terms of the energy density of different fluids comprising the matter sector is given by ${\cal{H}_{\mathrm{matt}}} = \sum_i \rho_i V$. For the non-interacting fluids, as assumed in this work, their energy densities $\rho_i$ satisfy the conservation law: 
\begin{equation}\label{cons}
\dot{\rho_i} + 3H(\rho_i + P_i) = 0.
\end{equation}
In the case of constant equation of state, $w_i=P_i/\rho_i$, where $P_i = -\frac{\partial\mathcal{H}_{\text{matt}(i)}}{\partial V}$ denotes the pressure,  the conservation law can be used to solve for $\rho_i$ such that, 
\begin{equation}
\rho_i = \rho_{0,i} a^{-3(1+w_i)},
\end{equation} 
where $\rho_{0,i}$ corresponds to the initial condition of energy density. A sum over all instances of $\rho_i$ and $P_i$ leads to expressions for total $\rho$ and $P$.

As mentioned earlier, in the Hamiltonian constraint (\ref{hcons1}) we have neglected the inverse volume modifications. These enter at two places in the quantum Hamiltonian constraint (see for eg. \cite{ck-closed2}). First by multiplying the gravitational part of the constraint by a term $A(V) = (V + V_c - |V - V_c|)/(2 V_c)$ where $V_c = 2 \pi \gamma \lambda \lp^2$. Note that $A(V) \rightarrow 1$ for $V \geq V_c$. The second modification arises in ${\cal{H}}_{\mathrm{matt}}$ depending on the equation of state of the matter. The inverse volume factors in the matter Hamiltonian are replaced by the eigenvalues of the corresponding operator in the quantum theory. Since the energy density for a matter with equation of state $w$ varies as $\rho \propto a^{-3(1+w)}$, matter Hamiltonian for a massless scalar field which has an equation of state of a stiff fluid ($w=1$) has the strongest contribution from the inverse volume modifications in contrast to dust or radiation or any other matter satisfying $w < 1$. In the quantum Hamiltonian constraint for the massless scalar, 
these modifications enter by multiplication of $g(V)^{
12}$ where $g(V) =  (V^{1/3}/V_c) (\sqrt{V+V_c} - \sqrt{|V-V_c|})$ \cite{ck-closed2}. 
The term $g^{12}$ is quickly approximated by $1/V^2$ for volumes larger than $V_c$. As an example, at volumes probing the Planck regime, say $V= 10$ Planck volume, their difference is approximately $0.2 \%$, and at $V = 100$ Planck volume the difference is approximately $2 \times 10^{-7}$. The effect of inverse volume terms is thus negligible when dynamics is probed at volumes larger than the Planck volume. It is to be noted that the inverse volume modifications affect the Hamiltonian constraints in the holonomy and the connection quantization in the same way. In the effective spacetime description, these modifications are difficult to implement without further assumptions. The reason is that these modifications which are inherently discrete are significant only close or below the Planck volume where the effective spacetime continuum itself becomes questionable. The following analysis in the effective dynamics will be based on the caveat of ignoring these modifications. (Their potential effect in the deep 
Planck regime will be discussed in Sec. V).

Using Hamilton's equations in the effective Hamiltonian constraint (\ref{hcons1}), the equation of motion for $V$ becomes

\begin{equation}
\dot{V} = \{V,\mathcal{H}_{\mathrm{eff}}\}   = - 4 \pi G \gamma \frac{\partial \mathcal{H}_{\mathrm{eff}}}{\partial \beta} = \frac{3}{\gamma \lambda}V \sin(\lambda \beta - D)\cos(\lambda \beta - D).
\end{equation}

A straightforward exercise involving the above equation and using the vanishing of the Hamiltonian constraint ${\cal H}_{\mathrm{eff}}^{\rm{(Hol)}} \approx 0$ yields the 
modified Friedmann equation,
\begin{equation}\label{modfried1}
H^2 = \frac{\theta^2}{9} = \frac{8\pi G}{3} (\rho - \rho_1 ) \left( 1-\frac{\rho - \rho_1}{\rcr}\right)
\end{equation}
where $\rcr = 3/(8 \pi G \gamma^2 \lambda^2) \approx 0.41 \rho_{\rm{Pl}}$ is the maximum allowed energy density in the case of the spatially flat FLRW model in LQC \cite{slqc}. And, 
\begin{equation}\label{rho1}
\rho_1 = \rcr[(1+\gamma^2)D^2 - \sin^2 D]. 
\end{equation} 
Note that unlike the bounce density in the spatially flat model, the bounce density in the $k=1$ model for the holonomy case has no bound at all because of the presence of $D$ term which is inversely proportional to $V^{1/3}$. Only in the case when inverse volume modifications are present, the energy density is bounded \cite{ck-closed2}. 
In eq.(\ref{modfried1}), $\theta$ is the expansion scalar which is universally bounded \cite{cs-geom}: $|\theta| \le 3/2\lambda\gamma$. Note that the Hubble rate vanishes at 
$\rho = \rho_1$ where the classical recollapse occurs for matter satisfying strong energy condition, and at $\rho = \rho_1 + \rcr$ where the classical big bang/crunch singularity is resolved and quantum bounce occurs. The quantum bounce is a direct manifestation of the underlying quantum geometry encoded via the parameter $\lambda$. Note that for phantom matter there is no big bang/crunch singularity, but a future big rip singularity. In this case, the bounce is classical and the recollapse is quantum. In the case when $\lambda$ is vanishing, $\rcr$ becomes infinite and the singularity resolution disappears.

One can similarly obtain the time variation of $\beta$ using Hamilton's equation, which yields:
\begin{equation}
\dot{\beta} = -4\pi G \gamma (\rho - \rho_2 + P) 
\end{equation}
where $\rho_2$ is given by,
\begin{equation}\label{rho2}
\rho_2 = \frac{\rcr D}{3}[2(1+\gamma^2)D - \sin(2\lambda \beta - 2D)-\sin(2D)].
\end{equation}
Finally, the time derivative of the Hubble rate at the quantum turnaround is given by
\be\label{dotH_hol}
\dot H = -\frac{1}{3 \gamma} \dot \beta ~.
\ee
Note that the nature of the quantum turnaround, whether it is a bounce or a recollapse, 
is determined entirely by the sign of $\dot \beta$. As we will see, this changes in the connection quantization with some interesting 
implications.

The differential equation in $V$, together with the differential equation in $\beta$ can be used to obtain the dynamical evolution. It is 
straightforward to check that these two equations are consistent with the conservation law for total energy density. Alternatively, the conservation law for the energy density along with either $\dot V$ and $\dot \beta$ can be used to obtain dynamical trajectories. In our analysis, we imposed initial conditions on 
$\rho$ and $V$,  and the initial condition on $\beta$ is found through the Hamiltonian constraint. This ensures that the effective Hamiltonian constraint is satisfied in numerical solutions. That is, 
\begin{equation}
\sin^2(\lambda \beta_0 - D_0) = \rho_0 \left(\frac{8 \pi G \gamma^2 \lambda^2}{3} \right) + \sin^2(D_0) - (1 + \gamma^2) D_0^2,
\end{equation} 
where subscript `0' denotes value at initial condition. Results from the numerical simulations based on these dynamical equations will be discussed in Sec. III.

\subsection{Connection operator quantization}
In the connection operator approach to the quantization of $k=1$ FLRW spacetime, instead of considering holonomies of the connection over closed loops one constructs a connection operator using open holonomies \cite{ck-closed1}. The resulting quantum theory is inequivalent to the one discussed above. The quantum Hamiltonian constraint is again a quantum difference equation with uniform steps in volume and is non-singular, but with some key differences in comparison to the holonomy quantization pertaining to the way quantum geometry effects influence spatial curvature. 
The effective Hamiltonian for this quantization is \cite{ck-closed1}:
\be\label{hcons2}
\mathcal{H}_{\text{eff}}^{\text{(con)}} = \frac{-3}{8 \pi G \gamma^2 \lambda^2} V [ (\sin(\lambda \beta) - D)^2 + \gamma^2 D^2] + {\cal H}_{\text{matt}} ~.
\ee
Here as in the holonomy quantization case, we have neglected the inverse volume modifications. As discussed in the previous subsection (below eq.(2.5)), these modifications turn out to be exactly the same as in the holonomy quantization. That is, the Hamiltonian constraint is modified in the identical way. Their effect  is negligible in comparison to the modifications originating from the trigonometric terms in the above Hamiltonian constraint.

Using Hamilton's equations,  we can obtain the modified Friedmann equation and the expansion scalar:
\begin{eqnarray}
H^2 = \frac{\theta^2}{9} &=& \frac{1}{\gamma^2	 \lambda^2} \cos^2(\lambda \beta)[\sin(\lambda \beta)-D]^2 \\
&=& \frac{8 \pi G}{3} (\rho - \rho_3 ) \left( 1-\frac{\rho - \rho_4}{\rcr}\right)
\end{eqnarray}
where $D$ is given by eq.(\ref{Deq}). Further,
\be\label{rho34}
\rho_3 = \gamma^2 D^2 \rcr, ~~~~~~ \mathrm{and} ~~~~~  \rho_4 = D((1 + \gamma^2)D - 2\sin \lambda \beta) \rcr ~.
\ee
In contrast to the holonomy quantization, $\rho_3$ is the energy density at the classical 
recollapse for matter satisfying strong energy conditions. At the quantum turnaround,
\be\label{con-rho}
\rho = \rcr(D^2(1 + \gamma^2) - 2 \sin \lambda \beta D + 1) ~.
\ee
Before going further, let us note some crucial differences between the holonomy and the connection quantization which are apparent so far. First, note that 
since $D$ is not bounded, the expansion scalar in the connection quantization is not bounded in contrast to the universally bounded behavior of $\theta$ in holonomy quantization. Second, the energy density at classical recollapse in connection quantization agrees with the classical expression, whereas $\rho_1$ in eq.(\ref{modfried1}) agrees only in the large volume limit. Finally, we see from eq.(\ref{con-rho}) that unlike the holonomy quantization there is not one but two values of energy density at which quantum turnaround can occur. Note that a quantum turnaround occurs if $\sin(\lambda \beta) = \pm 1$. The two energy densities at the quantum turnaround then correspond to
\be\label{conrho-bounce}
 {{\rho^\mp}} = \rcr((D \mp 1)^2 + \gamma^2 D^2) ~, 
\ee
respectively for positive and negative unit values of $\sin(\lambda \beta)$.

The Hamilton's equation for $\beta$ yields
\begin{equation}
\dot{\beta} = -4 \pi G \gamma(\rho - \rho_5 + P),
\end{equation}
where $\rho_5$ is given by, 
\begin{equation}
\rho_5 = \frac{2 \rho_{\text{max}}D}{3}[(1+\gamma^2)D - \sin(\lambda \beta)].
\end{equation}
Using above equations, the time rate of change of Hubble rate at the quantum turnarounds can be obtained as
\be \label{dotH_conn}
\dot H = -\dot \beta (1 \mp D)
\ee
where the negative (positive) sign corresponds to the quantum turnaround occurring due to $\rho_4^-$ $(\rho_4^+)$. This is interesting when compared to the 
time rate change of Hubble rate in the holonomy quantization. In the latter case, one quantum turnaround is identified as either bounce or 
recollapse (in the phantom case), but its nature is explicitly determined by the value of $D$. In the connection case, quantum turnarounds 
are of two types and they depend explicitly on $D$. As we will discuss later, for the case when $D$ becomes extremely 
small, both the quantum turnarounds become less distinct.

Along with $\dot V$ and the conservation law (\ref{cons}), above equation forms a consistent set of dynamical equations of which any two can be used to obtain dynamical solutions. In the numerical simulations, $V_0$ and $\rho_0$ are imposed and $\beta_0$ is solved for from the effective Hamiltonian,
\begin{equation}
\sin(\lambda \beta_0) = \left[\rho_0 \left(\frac{8 \pi G \gamma^2 \lambda^2}{3} \right) - \gamma^2 D_0^2\right]^{1/2} + D_0.
\end{equation}

It is important to note that the initial values of $\beta_0$ for the holonomy and connection quantization will differ depending on the initial conditions of $V_0$ and $\rho_0$. This can give rise to phenomenological differences for the solutions, but as long as initial conditions are ``classical" (i.e. large volume and low energy density) the difference in $\beta$ is quite small.  

Finally, it is interesting to note that the energy density at the quantum bounce is generally larger in the connection quantization than in the holonomy case if the bounce occurs at same volume in both the cases. To see this, we can compare  the energy density at quantum turnaround in holonomy case $\rcr + \rho_1$ (see eq.(\ref{rho1})), and the energy densities for distinct quantum turnarounds in the connection case $\rho^\pm$ given by eq.(\ref{conrho-bounce}). It is straightforward to see that $\rho^+$ is always greater than $\rcr + \rho_1$. On the other hand, $\rho^-$ can be less than $\rcr + \rho_1$ if the bounce volume is such that $D > 1/2$. Using eq.(\ref{Deq}) this implies, that if the bounce volume $V_b$ is  is such that $V_b < 16 \pi^2 \lambda^3$, then $\rho^-$ at that bounce volume can be less than  $\rcr + \rho_1$ at that same bounce volume. In all of the other situations, we find that the density at quantum turnaround  in holonomy quantization is always less than the densities $\rho^\pm$ in the 
connection 
quantization if they are compared at the same bounce volume. It is also interesting to note that the terms in $\rho^\pm$ which are responsible for causing energy density at the bounce volume to be generally larger than $\rcr + \rho_1$, originate from the terms which cause unboundedness of the expansion scalar in the connection quantization. In this particular way, one may view that lower energy density at the bounce, if it occurs at the same volume such that $V_b > 16 \pi^2 \lambda^3$,  in the holonomy quantization in comparison to the connection quantization and the respective behaviors of expansion scalars in the two quantization prescriptions are related. However, depending on the initial conditions and the matter content,  the bounce volumes in the connection and the holonomy cases are different. And cases with lower bounce volumes have higher energy density at the bounce. In our numerical simulations, we found that generically the holonomy bounce is always at a higher volume than at least one of the two 
distinct bounces 
in the connection quantization. Therefore, the energy density at the bounce in the holonomy case is always lower than the bounce density at one of the distinct quantum bounces in the connection case.

\section{Bounces in the Connection and the Holonomy Quantization: fluid(s) satisfying strong energy condition}
In this section, we consider the solutions obtained from numerically solving the dynamical equations in the connection and the holonomy quantization approaches for the single fluid and the two fluid cases. The considered fluids satisfy strong energy condition. In particular, for the single fluid case we consider equations of state $w=1$, $w=1/3$, $w=0$ and $w = -0.3$. We start with a brief discussion of numerical accuracy, followed by the results in 
single fluid cases, and two fluid scenarios.

\subsection{Numerical Accuracy}
In the numerical simulations considered in this manuscript, we use an adaptive scheme to solve all of the coupled differential equations. 
We report good error control in all solutions presented. Because the Hamiltonian constraints $\mathcal{H}_{\rm{eff}}^{\rm{(hol)}}$ given by 
eq.(\ref{hcons1}) and $\mathcal{H}_{\rm{eff}}^{\rm{(con)}}$ given by eq.(\ref{hcons2}) vanish, numerically we must obtain these to be 
approximately equal to zero at all times. This vanishing of the Hamiltonian constraint in numerics can be used to check the validity of our 
solutions. Plotted in Fig. 2  is the relative error for two representative solutions from the connection and the holonomy quantizations. These 
correspond to the case of $w=1$ which is also discussed in Fig. 3 for independent simulations.

\begin{figure}[tbh!]
\includegraphics[scale=0.55]{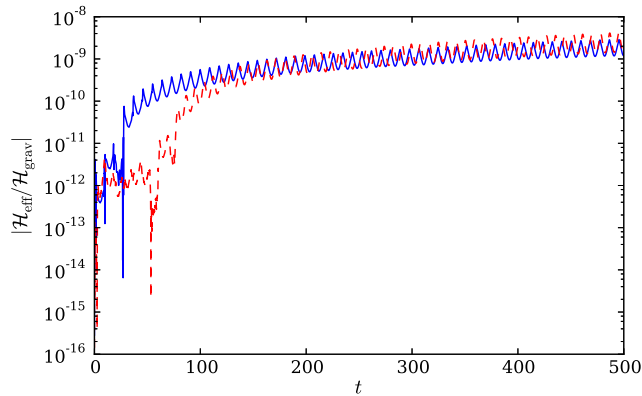}
\caption{Relative error in the effective Hamiltonian is plotted for representative holonomy and connection solutions. The solid curve corresponds to the holonomy solution, whereas the dashed curve represents connection solution. The initial conditions (in the Planck units) for this simulation are $V_0 =1000$, $\rho_0 = 0.2$, with $\beta_0$ set from the Hamiltonian constraint. The equation of state is $w = 1$. }
\end{figure}

Since we are dealing with multiple bounces, due to the cyclic nature of spatially closed universes in LQC, and that the values of derivatives of the scale factor at the bounce can be very high, it is important to check that errors remain low after successive bounces. One may notice from Fig. 2 that the error increases in time. It is very important to check that the errors do not grow exceedingly large on the time scale of interest. In this particular case this issue is not significant because the errors only increase to the scale of $10^{-8}$. But in some other cases we found that errors can grow exponentially in a relatively short time scale. To ensure this is not the case, we checked the error for every solution presented in this paper. The errors involved in the forthcoming solutions discussed in this manuscript are on or around (within 1 or 2 orders of magnitude) the scales displayed by Fig. 2. Note that a plot of $|\mathcal{H}_{\rm eff}/\mathcal{H}_{\rm matt}|$ produces essentially the same error as Fig.
 2.

\subsection{Single Fluid Universes}

A crucial result of connection quantization reported earlier by Corichi and Karami  is that of existence of two 
distinct bounces characterized by the different values of the volume for the massless scalar field ($w=1$) \cite{ck-closed1}. We start with 
a repeat of those findings here. In our numerical investigations we find that for $w=1$, the result is independent of initial conditions. 
Figs. 2 and 3 show such cases.

\begin{figure}[tbh!]
\centering
\includegraphics[scale=0.55]{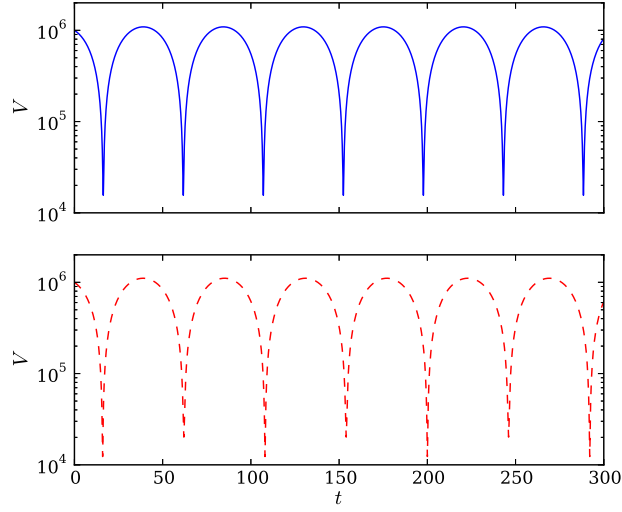}
\caption{Variation of volume versus time for a massless scalar field ($w=1$) is shown. The connection quantization is represented by dashed 
line (bottom panel) while the holonomy quantization is represented by the solid line (top panel). The initial conditions for this particular 
solution were as follows: $V_0 = 10^6,\rho_0 = 10^{-4}$ (in Planck units), with $\beta_0$ set by the Hamiltonian constraint. The energy 
density at the bounce in the holonomy case is $\rho = 0.411 \rho_{\rm{Pl}}$. In the connection case, distinct quantum bounces occur at $\rho = 0.658 \rho_{\rm{Pl}}$  and  $\rho = 0.247 \rho_{\rm{Pl}}$. }
\end{figure}

It is important to note a few key things about Fig. 3. First, as mentioned in the previous section, the initial condition on $\beta_0$ 
is different for each method of quantization because of the different functional forms of each respective Hamiltonian constraint. $\beta_0$ 
has typical values of the order of $10^{-2}$ for initial conditions similar to those selected for Fig. 2. Such a choice of initial 
conditions corresponds to classical universe initial conditions because they are given when GR is a very good approximation to LQC. In Fig. 4, we show a zoom of the initial evolution starting from the GR regime for the same initial conditions as in the case of Fig. 3. In the early period, the GR solution and the LQC solution (shown only for the holonomy case for clarity) match but after some time they depart from each other as the singularity is approached. We find that the GR solution results in a 
big crunch singularity in a finite time where the volume of the universe vanishes. On the other hand, the LQC universe undergoes a quantum bounce, resolving the singularity. In Fig. 4, we thus see two important contrasting features: (i) the way GR description departs from LQC as the singularity is approached, and (ii) the cyclic behavior in LQC contrasted with the one in GR. 

\begin{figure}[tbh!]
\centering
\includegraphics[scale=0.5]{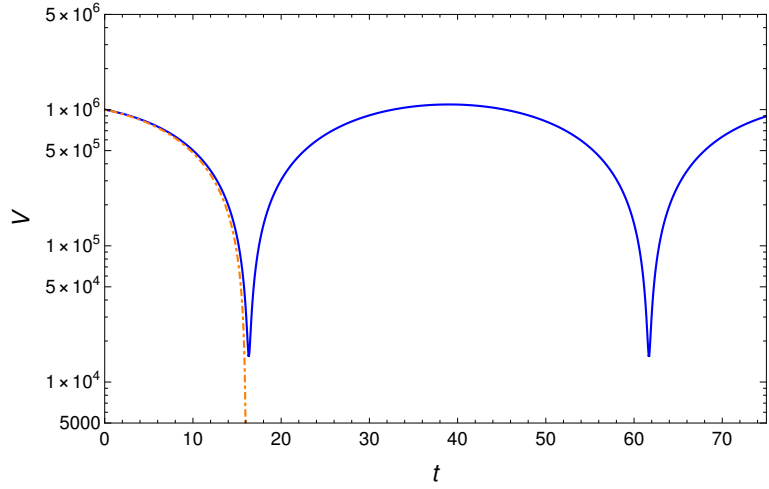}
\caption{This plot shows the comparison of GR evolution with holonomy case in LQC for the early period corresponding to Fig. 3. For clarity, we only show the holonomy case. The connection case yields a very similar picture of the contrast between GR and the LQC. Solid curve corresponds to the LQC evolution and dashed curve depicts the evolution in GR. After agreement initially, two curves show departures. The GR curve ends in a big crunch singularity, whereas the LQC curve undergoes cyclic evolution. Bounces in the LQC evolution occur at $\rho = 0.411 \rho_{\mathrm{Pl}}$. }
\end{figure}

In 
contrast to the classical universe initial conditions, for a quantum universe initial conditions, difference in values of $\beta_0$ are of the order of $10^{-2}$ and $10^{-1}$, with 
$V_0$ and $\rho_0$ of the order of $10^3$ and $10^{-1}$ (in Planck units). (With quantum universe initial conditions we imply those 
universes which do not have a classical epoch in the evolution). For the simulations in Fig. 3, the values of $\beta_0$ are: 
$\beta_0^{\rm{(hol)}} = 0.0247580330047$ and $\beta_0^{\rm{(con)}} 
= 0.0245774931055$. The relative difference being approximately $10^{-2}$ results in a very similar evolution for the holonomy and the connection 
quantizations at early times as is evident from Fig. 3. On the other hand, for the simulations in Fig. 2, $\beta_0^{\rm{(hol)}} = 
-0.0407652373014$ and $\beta_0^{\rm{(con)}} = -0.030416331822$, yielding a larger relative difference. 
The relative difference in $\beta_0$ is magnified for quantum universe initial conditions,
 and it is usually the case that the methods disagree. (Such cases are discussed in the subsequent sections). 
Note that the differing values of $\beta_0$ for each 
quantization method is present throughout all of this work. It is a necessity because it is the only way one can ensure that the Hamiltonian constraint is satisfied. 

In Fig. 3, the two distinct types of bounces in connection quantization can be clearly contrasted with only one type of bounce in the holonomy quantization. 
The demonstrated simulations begin in a contracting universe. The scale factor is forced downward until a bounce occurs and is followed by a period of expansion 
until turnaround occurs due to positive curvature. Instead of repeating a cyclic behavior as in the holonomy quantization, the connection quantization bounces at different values of $V$ in subsequent cycles, with distinct bounces alternating each other. The connection quantization probes deeper volumes than the holonomy quantization every other bounce and correspondingly probes energy densities that are about 1.5$\rcr$. In contrast,  the holonomy quantization has a bounce occurring very near to $\rcr$ every time a bounce occurs. Interestingly, the bounces at the higher volume in the connection quantization (e.g. the bounces at $t\approx 60,\; 150,\; 240$)  occur at an energy density of about 0.60$\rcr$.

For the simulations shown in Fig. 3 and 4, the effect of inverse volume modifications is negligible. Since the bounces occur at volumes greater than $V_c$, only the matter part of the Hamiltonian constraint gets affected. For the holonomy quantization, the maximum relative difference in the energy density computed with and without inverse volume modifications turns out to be approximately $4.5 \times 10^{-12}$. (The maximum relative difference is computed at the bounce volume). On the other hand, for the connection quantization the maximum relative difference is approximately $9.2 \times 10^{-11}$. For the case of Fig. 2, bounces occur at lower volumes, approximately $665 ~ V_{\mathrm{Pl}}$ in the holonomy quantization and approximately $375 ~ V_{\mathrm{Pl}}$ and $ 1466 ~ V_{\mathrm{Pl}}$. The maximum relative error is approximately $5.9 \times 10^{-8}$ for the holonomy quantization and approximately $3.3 \times 10^{-7}$ for the connection quantization. The small difference in the holonomy and connection 
quantization occurs because of the lower bounce volume in the connection quantization. For other simulations discussed in this section, we find similar negligible differences for ignoring the inverse volume modification.

The next three figures show that the two bounce phenomena holds true for the radiation-dominated ($w=1/3$), matter-dominated ($w=0$), and $w=-0.3$ universes.  
For all these equations of state, strong energy condition is satisfied and inflation does not occur. After the decelerated expansion leading to a classical recollapse, quantum gravitational effects in holonomy and connection quantizations result in avoidance of big crunch and multiple cycles of contraction and expansion.

\begin{figure}[tbh!]
\centering
\includegraphics[scale=0.55]{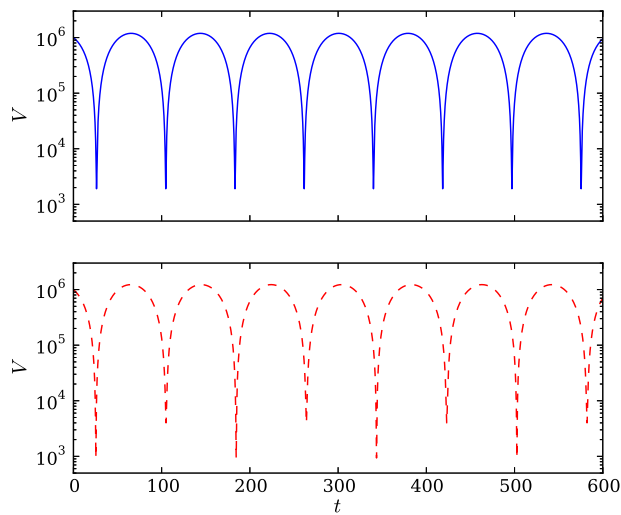}
\caption{Behavior of volume of the universe versus time is shown for a radiation-dominated ($w=1/3$) universe. Initial conditions for volume and energy density, and designations for these solutions are the same as in Fig. 3. The energy density at the bounce in the holonomy case is $\rho = 0.423 \rho_{\rm{Pl}}$. In the connection quantization, the distinct bounces occur at $\rho = 1.095 \rho_{\rm{Pl}}$ and at $\rho = 0.157 \rho_{\rm{Pl}}$.}
\end{figure}

\begin{figure}[tbh!]
\centering
\includegraphics[scale=0.55]{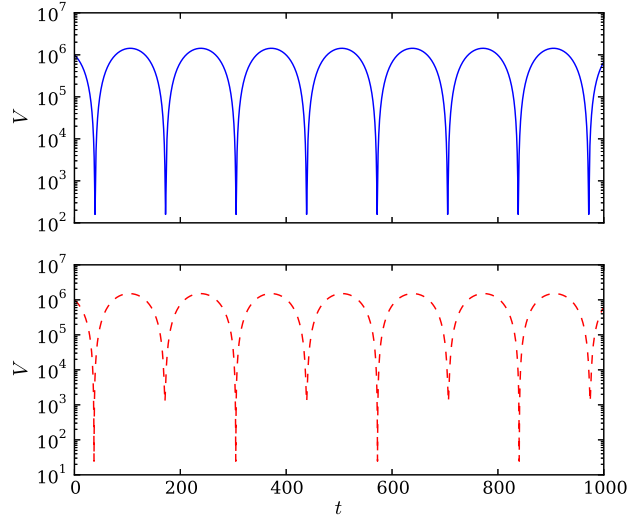}
\caption{$V$ vs. $t$ for a matter-dominated ($w=0$) universe is demonstrated. The choice of $V_0$ and $\rho_0$, and labeling of solutions is same as in Fig 3. 
The bounce in holonomy case occurs at $\rho = 0.629 \rho_{\rm{Pl}}$, and the distinct bounces in connection case occur at $\rho = 4.073 \rho_{\rm{Pl}}$ and at $\rho = 0.0817 \rho_{\rm{Pl}}$.} 
\end{figure}

\begin{figure}[tbh!]
\centering
\includegraphics[scale=0.55]{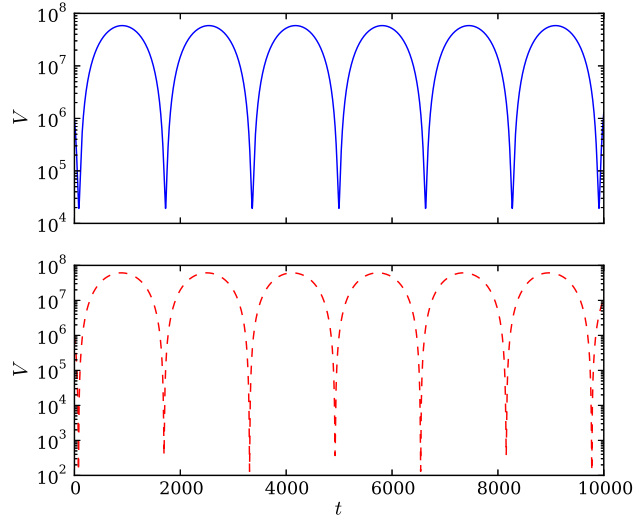}
\caption{Volume of the universe is plotted versus time for a universe with equation of state $w=-0.3$. Initial conditions and designations for these solutions are the same as in Figure 3. The energy density at the bounce in the holonomy case is  $\rho = 1.58 \times 10^{-3} \rho_{\rm{Pl}}$. In the connection case, the distinct bounces occur at $\rho = 0.0525 \rho_{\rm{Pl}}$ and $\rho = 0.0253 \rho_{\rm{Pl}}$. In contrast to simulations in previous figures, the energy density at bounce is smaller because for $w=-0.3$, $\rho \propto a^{-2.1}$ which is only slightly stronger than the way classical spatial curvature behaves ($a^{-2}$). The quantum turnarounds hence get a significant help from the spatial curvature term, reducing the role of matter density in comparison to previous cases in causing the bounce. }
\end{figure}

Of these examples, the most exaggerated case of the two bounce phenomena in the connection quantization is seen in the matter-dominated universe in Fig. 6. In this universe, the two bounces in the connection quantization are separated by about 2 orders of magnitude in volume!  The difference in the time between the bounces also gets increasingly larger as we decrease $w$. This can be explained in terms of the behavior of energy density. Because $\rho_{\rm Fig 2} \propto a^{-6}$ and $\rho_{\rm Fig 5-7} \propto a^{-4}, \; a^{-3}, \; a^{-2.1}$ respectively, it takes longer for positive curvature to overcome outward expansion in the case of $w=1/3, \;0, \; -0.3$. The case of $w=-0.3$ discussed in Fig. 7, is interesting due to one additional feature. In this particular case, the behavior of energy density with respect to the scale factor is very similar to the behavior of spatial curvature which scales as $1/a^2$. As the scale factor becomes small, the contribution due to the positive spatial curvature is 
significant in causing a turnaround from the to be classical singularity. Depending on the initial conditions, the quantum bounces thus can occur at lower energy densities than for more positive equations of state.

Results shown in Fig.3-7, are representative cases for a vast number of simulations we performed for single fluid scenarios for equation of state $w > -1/3$. 
Results in all these simulations confirm with the shown results -- two distinct bounces for connection quantization and one distinct bounce in holonomy 
quantization. In the case of a single fluid, we report the two bounce phenomena of the connection quantization to be a generic feature for $w>-1/3$. In the case when strong energy condition is violated, but $w \geq -1$, the spatially closed universe may undergo inflation and cycles do not occur. In case, inflation does not occur then the results are found to be in agreement to the ones discussed above.

\subsection{Two Fluid Universes}

\begin{figure}[tbh!]
\includegraphics[scale=0.55]{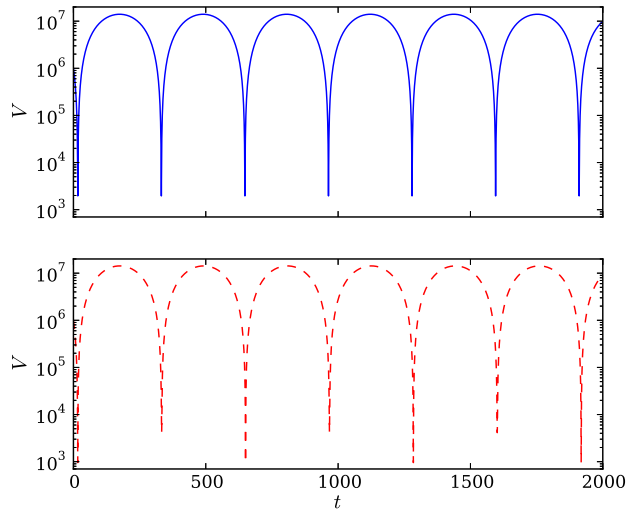}
\caption{$V$ vs. $t$ for a universe in a two fluid scenario is shown. The equations of state are $w_1=1/3$ and $w_2 = -0.2$.}
\end{figure}

If the universe is sourced with two non-interacting fluids, then the effective equation of state which governs the cosmological dynamics depends on the individual equations of state and the initial conditions for the respective energy densities. Using the effective dynamical equations for the holonomy and 
connection quantization we performed various simulations for different choices of equation of states for two fluid scenarios. Results agree with the case of 
single fluid scenario. 

When both the fluids have equations of state, $w_i > -1/3$, we find that in the holonomy quantization, effective dynamics results in a single distinct bounce 
over various cycles. For the initial conditions corresponding to the classical universe, the energy density at the bounce is always approximately equal to $\rcr$. In contrast, the effective dynamics of the connection quantization results in two distinct bounces in the cyclic behavior of the universe. 
An example of such a simulation is provided in Fig. 8. The figure corresponds to the case with $w_1 = 1/3$ and $w_2 = -0.2$. The initial conditions on volume and total energy density are same as in Fig. 3. In comparison to the simulation in Fig. 6, the two bounce phenomena is a bit subdued. It should be noted that the latter behavior is sensitive to the initial conditions. But, the two bounce phenomena is found to be always present irrespective of the initial conditions.

In case any or both of the equation of state(s) is $-1 \geq w_i \geq -1/3$, then depending on the initial conditions of the energy densities, there is a possibility for inflation to commence. If inflation does not occur, then the results are similar to above cases. However, if inflation occurs then the cyclic behavior does not 
take place and the two distinct bounce phenomena does not occur. The latter phenomena  occurs as long as one of the equations of state is  $w_i > -1.0$ and  $w_i$ does not lead to eternal inflation after a single bounce.


\section{Quantum recollapses in the Connection and Holonomy Quantization: phantom fluid(s)}
In classical GR, a universe sourced with matter with an equation of state $w < -1$ accelerates faster than in the case of the positive cosmological constant. For a fixed equation of state, the energy density increases as the scale factor increases, and decreases as the scale factor decreases. The latter leads to the avoidance of the big bang singularity in the classical theory. But a big rip singularity emerges in the future evolution. In LQC, phantom dynamics 
results in an interesting picture where following the bounce the phantom field captures the dynamics of a standard scalar field for some time \cite{ps6}. Earlier works in LQC, in the spatially flat model, show that this strong curvature singularity is resolved  due to the quantum geometric effects, and is replaced by a quantum recollapse \cite{ssd,lqc-bigrip3,*lqc-bigrip4}. For the spatially closed model quantized using closed holonomies, indications of the resolution of big rip singularity 
were found in Ref. \cite{psvt}. However, in all these studies, a detailed analysis of the phenomenological nature of singularity resolution in past and future for quantum universes and the two-fluid scenarios was so far not available. 

In the following we discuss the cosmological dynamics in the holonomy and the connection quantization for the case of a single fluid with a 
phantom equation of state and in a two fluid scenario. Interesting aspects of the dynamics appear depending on whether the initial conditions are chosen such that the universe is initially classical or the universe remains quantum in the evolution. These cases are discussed separately for the single fluid scenario.

\begin{figure}[tbh!]
\includegraphics[scale=0.38]{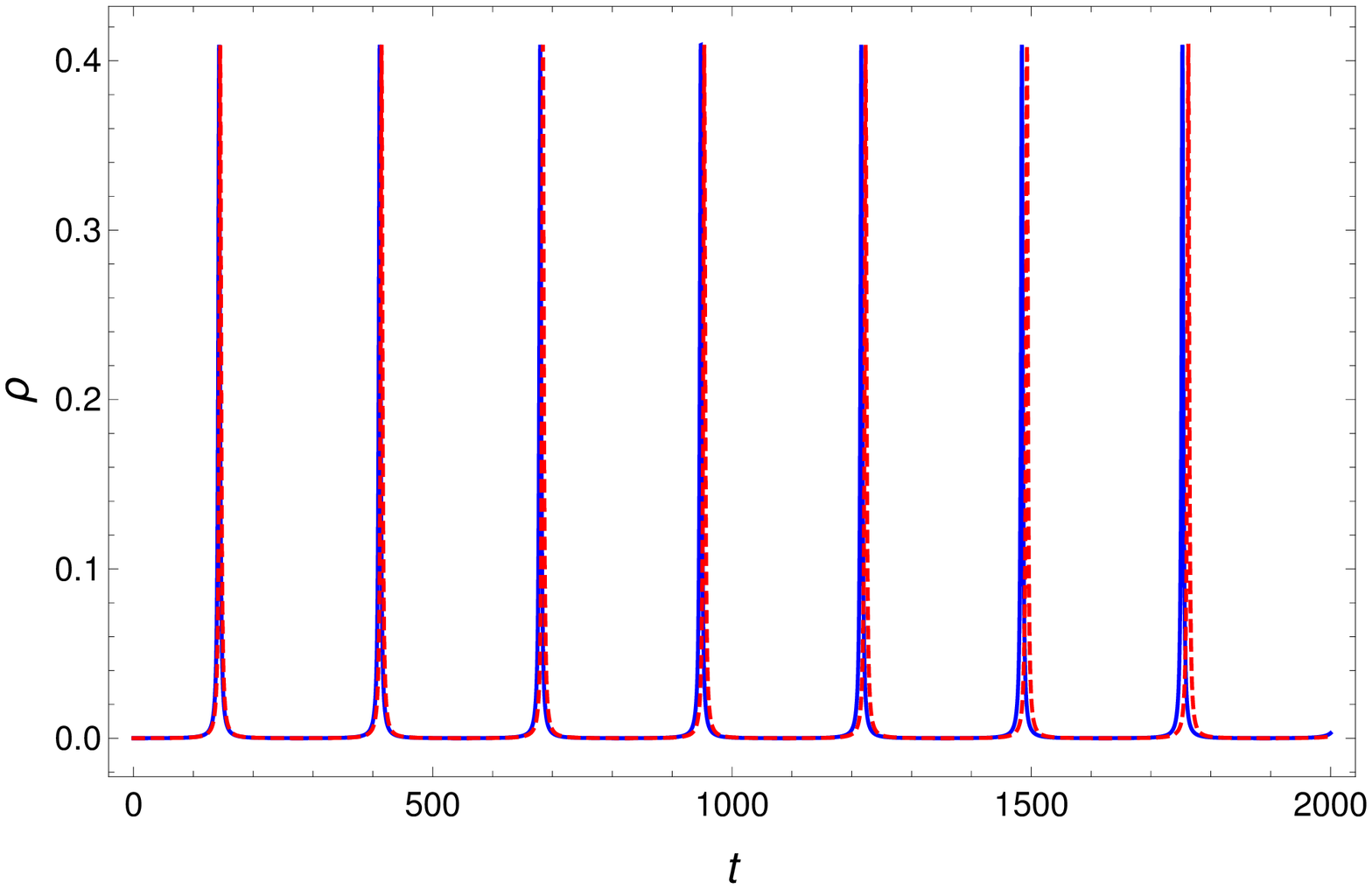}
\includegraphics[scale=0.38]{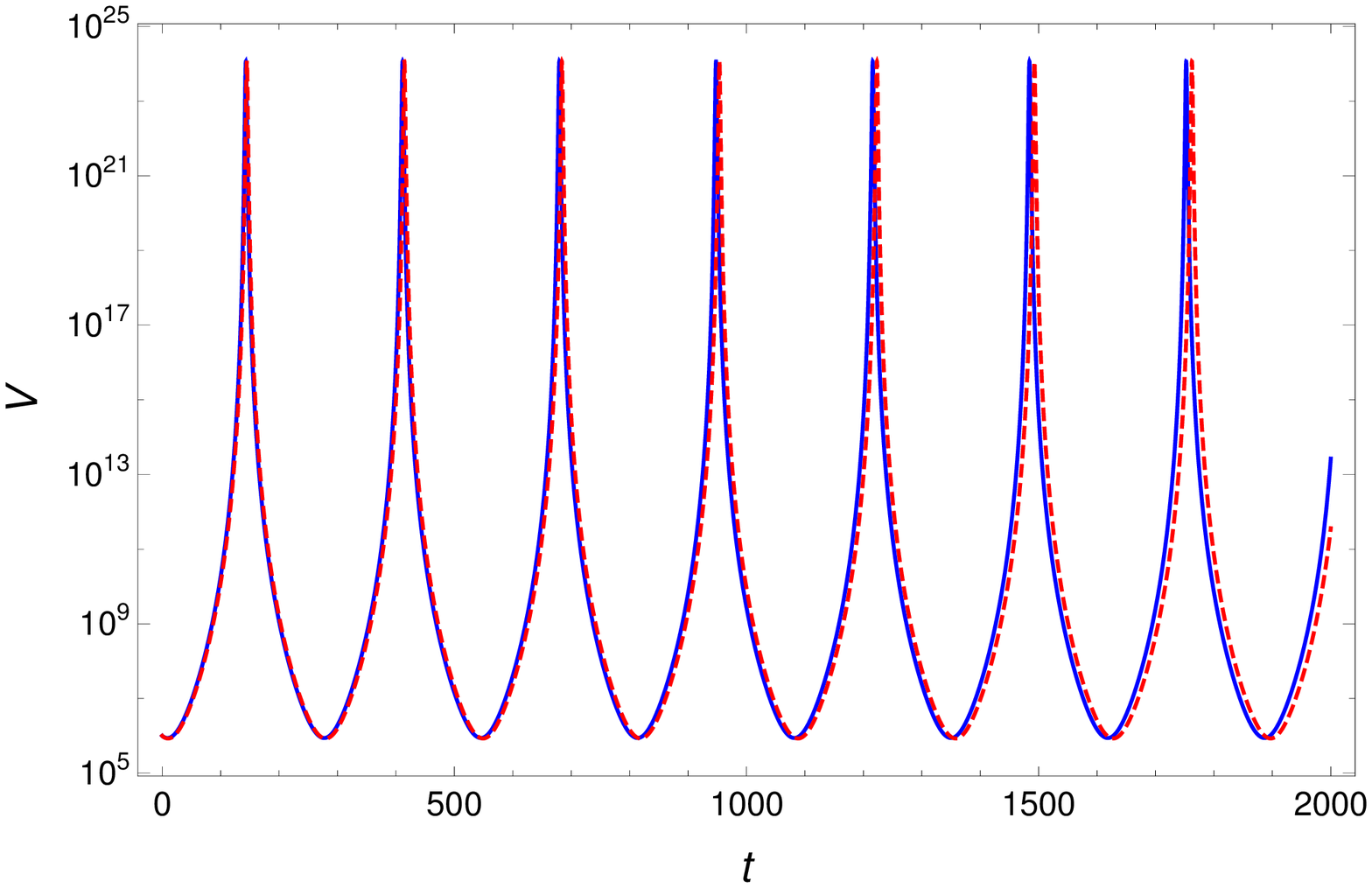}
\caption{Variation of energy density $\rho$ and volume $V$ with respect to time is shown for a universe with phantom energy of $w=-1.2$ in the connection quantization approach (dashed curve) and the holonomy quantization approach (solid curve). The initial conditions for this universe are $V_0 = 10^6$ and $\rho_0 = 10^{-4}$. The universe starts evolution from a classical regime and undergoes cyclic evolution with only a distinct quantum bounce in both the approaches.}

\end{figure}

\begin{figure}[tbh!]
\centering
\includegraphics[scale=0.38]{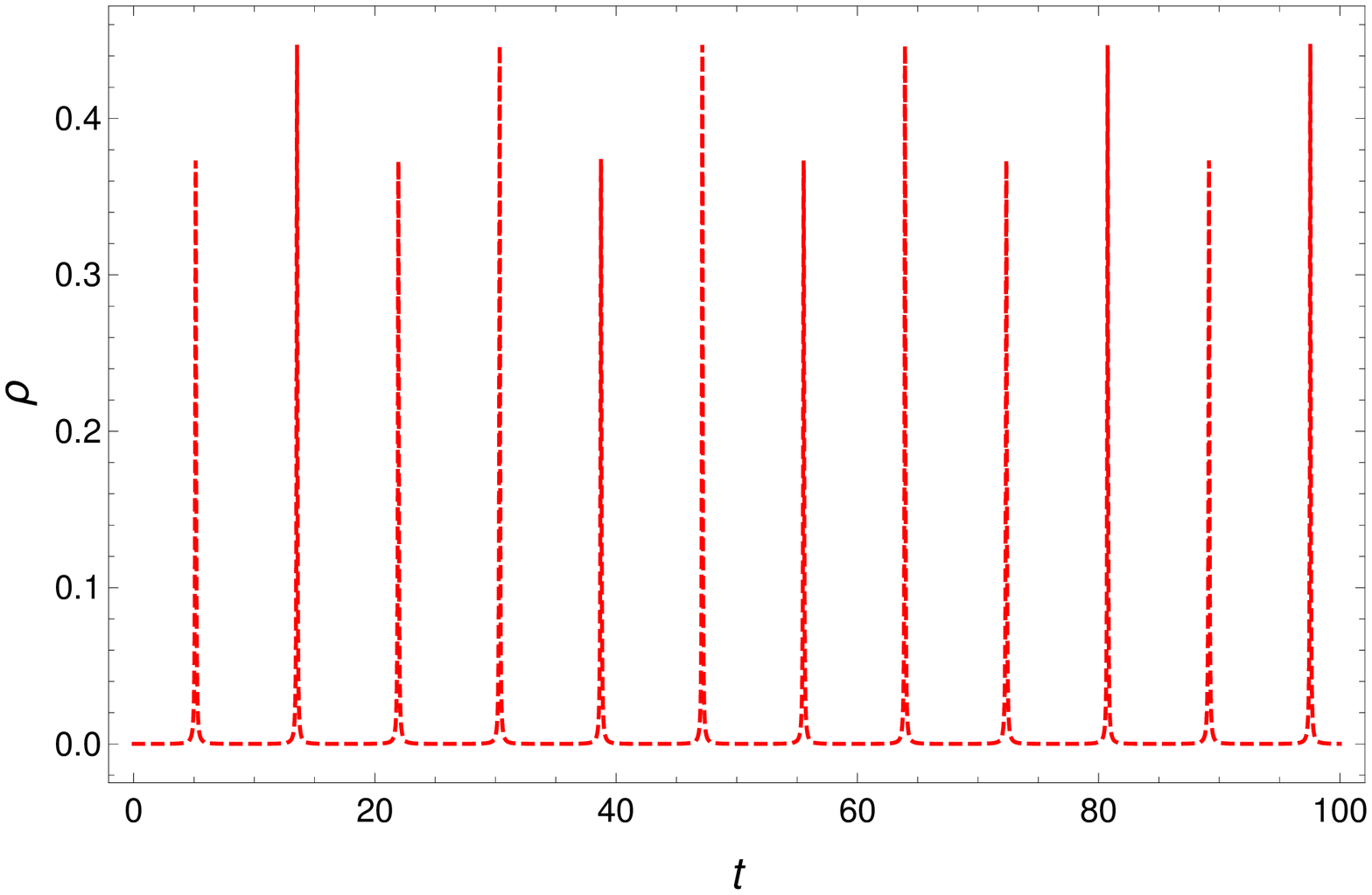}
\includegraphics[scale=0.38]{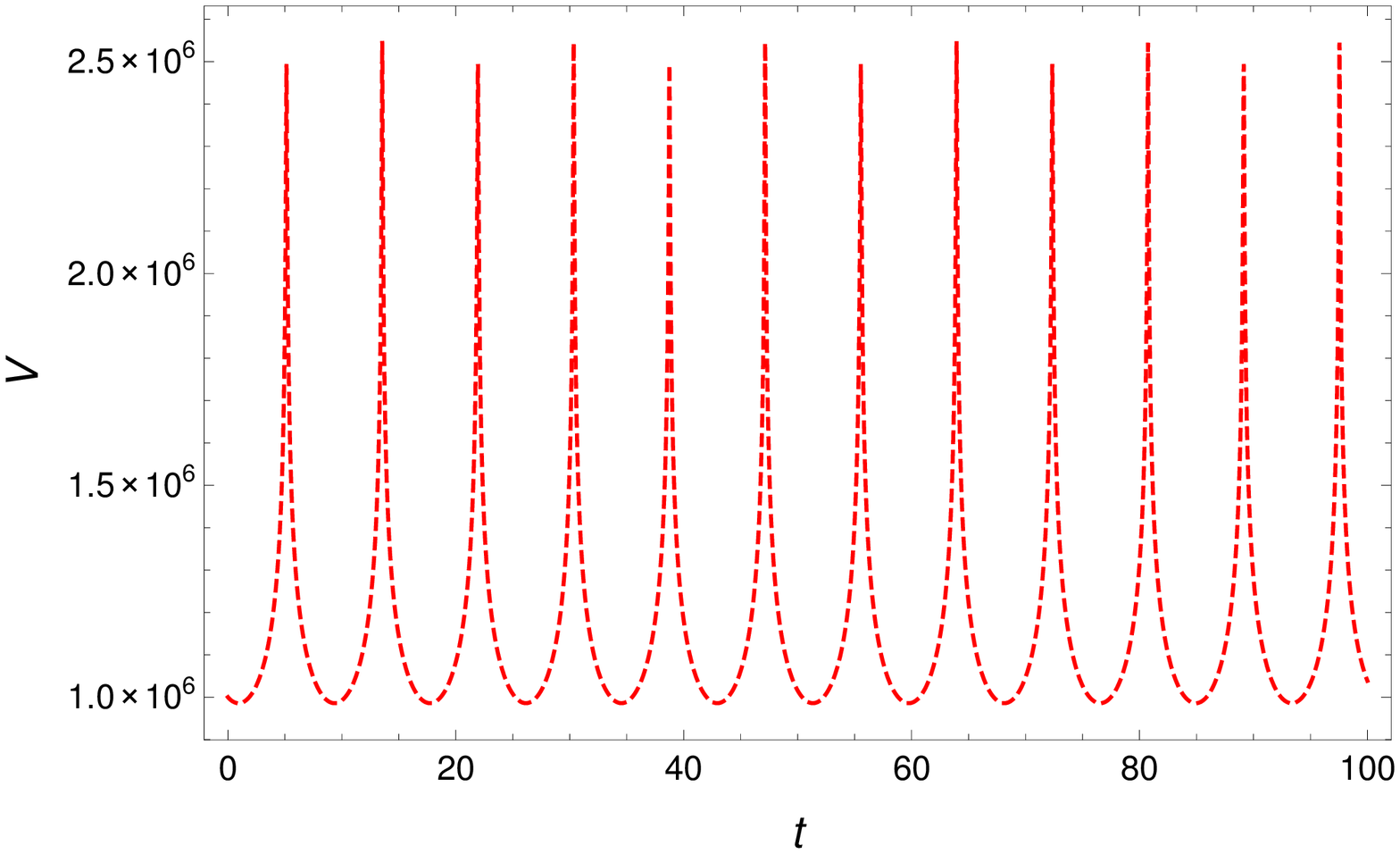}
\caption{Energy density and volume are plotted against time for connection quantization for the case of a universe with a strong phantom equation of state ($w = -10$). The universe starts from the initial conditions of $V_0 = 10^6$ and $\rho_0 = 10^{-4}$. Due to the strong phantom dynamics and the quantum geometric effects, avoidance of big rip appears at smaller volumes in comparison to Fig. 9 and therefore are distinct quantum recollapses.  }
\end{figure}

\subsection{Single Fluid Universes}
We first consider the case when the initial conditions are given such that the dynamics in LQC, whether connection or holonomy quantization, agrees very well with the dynamics of GR. Then we consider the case when initial conditions are given in the quantum regime such that the universe does not have a classical epoch.

\begin{figure}[tbh!]
\centering
\includegraphics[scale=0.48]{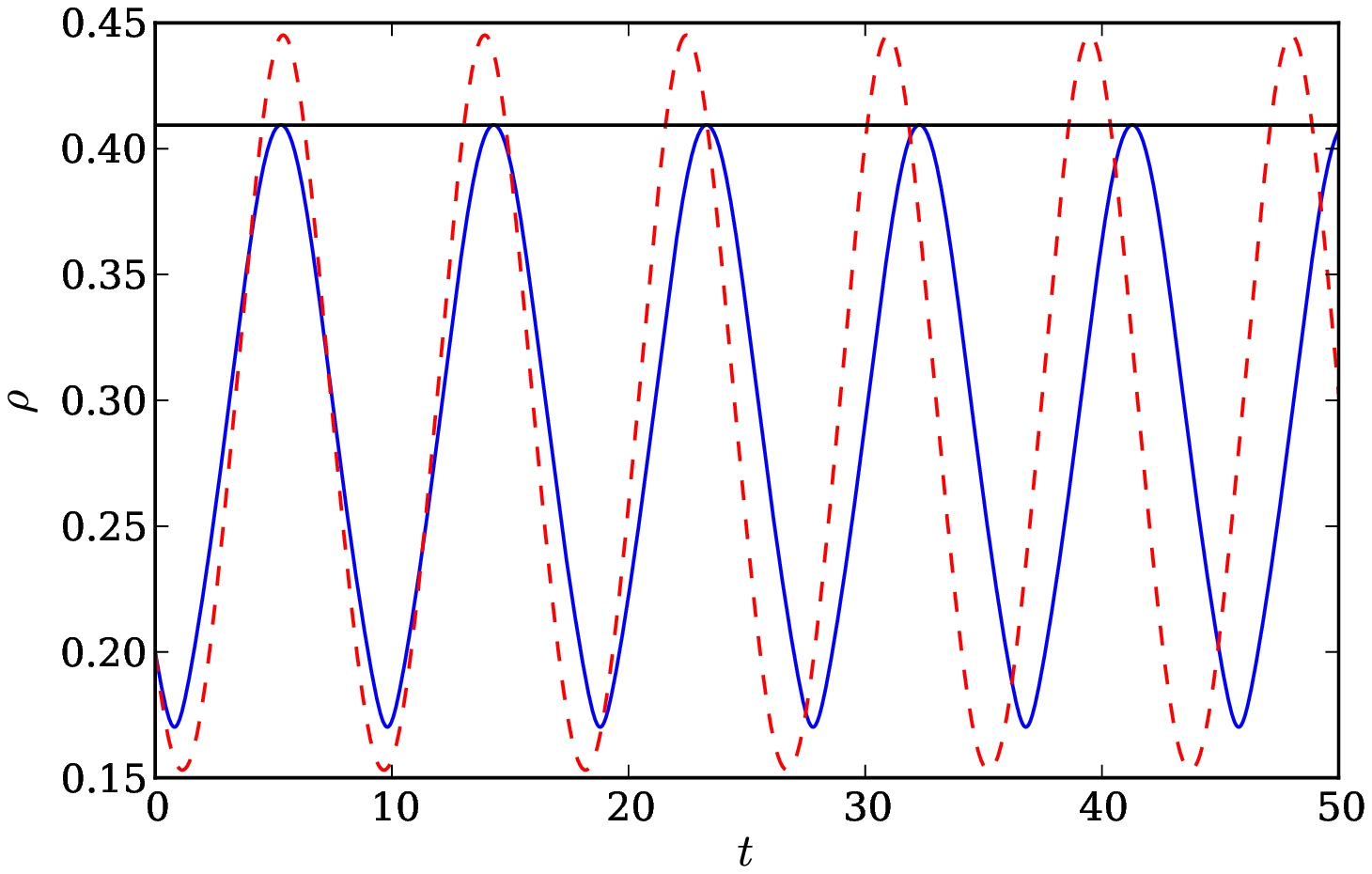}
\includegraphics[scale=0.48]{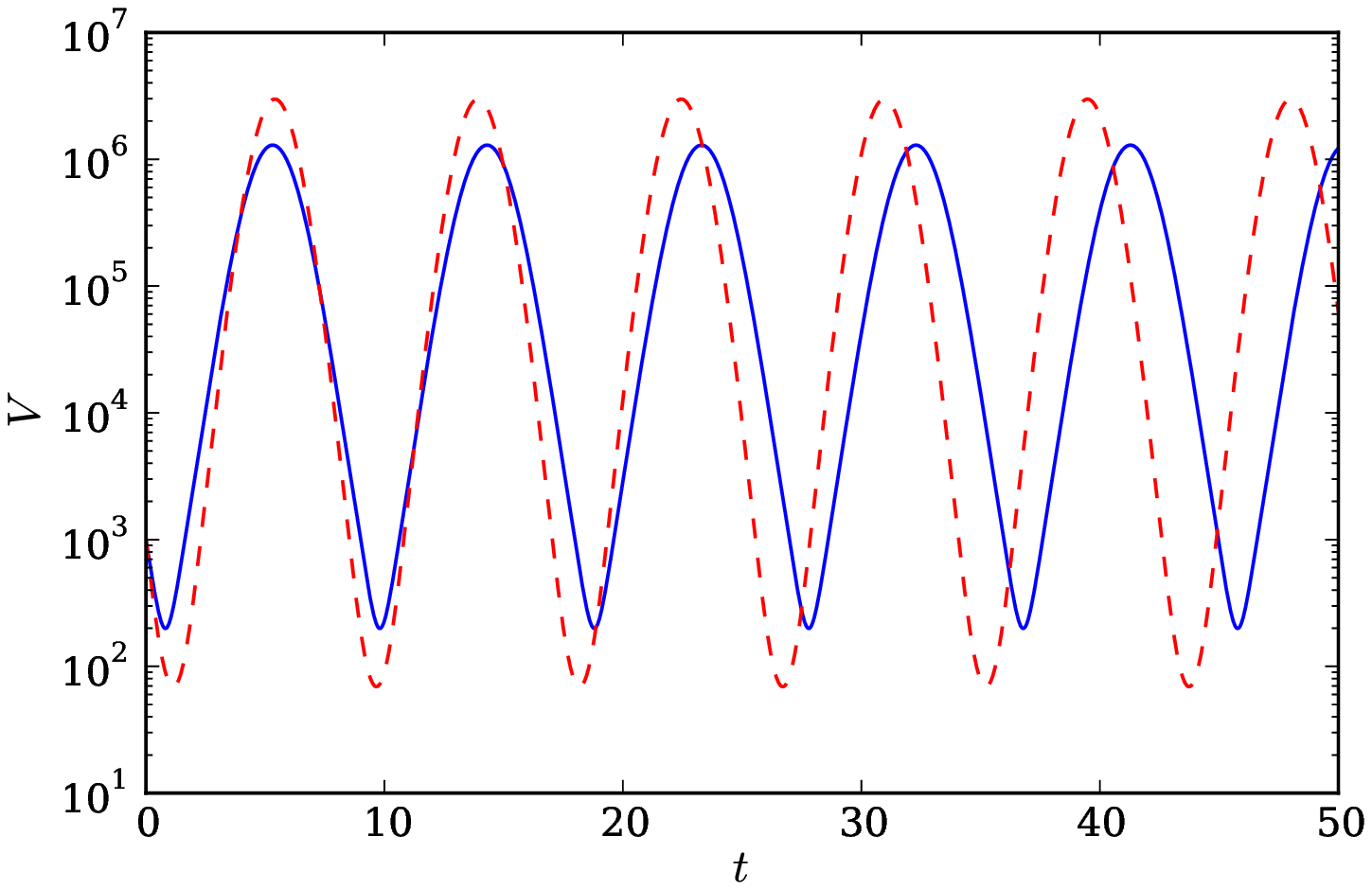}
\caption{The variation of $\rho$ and $V$ in time is shown for a universe with initial conditions $V_0 = 1000$ and $\rho_0 = 0.2$ and with a phantom equation of state, $w=-1.1$. The black horizontal line represents $\rcr$ and holonomy and connection designations are the same as in the earlier figures. Though the evolution in connection and holonomy quantization appears very similar there are important differences in nature of turnarounds as noted in the text. }
\end{figure}

\subsubsection{Classical universe initial conditions}
In the case when initial conditions are provided in the classical regime, dynamics in holonomy quantization results in a cyclic universe. A classical bounce at small volumes due to the phantom fluid and a quantum recollapse at large volumes when the energy density approaches $\rcr$. There is only a single distinct quantum turnaround. That the quantum turnaround is a recollapse can be checked by the sign of $\dot \beta$ which turns out to be positive, as a result of which $\dot H < 0$ (using eq.(\ref{dotH_hol})). In the connection quantization, the situation is more subtle. The universe is still cyclic, but unlike the case of fluids which satisfy the strong energy condition, there is only a single distinct quantum recollapse 
for a large class of initial conditions, and equations of state $w \lesssim -1$. However, distinct recollapses reappear albeit not generically when the phantom equation of state is such that $w \ll -1$. The reason for the lack of two distinct 	quantum recollapses is tied with them occurring at very large volumes. In the latter case, $D$ becomes vanishingly small and therefore $\rho^+_4$ and $\rho^-_4$ computed at the quantum turnaround (eq.(\ref{conrho-bounce})) become almost equal to each other.

In Fig. 9 we compare the dynamics for the phantom fluid in the holonomy and the connection quantizations. The equation of state is $w = -1.2$ with initial conditions corresponding to a classical contracting universe.  Such an equation of state  results in only one  distinct quantum recollapse followed successively by 
classical bounces both in holonomy and connection quantizations. Notably the energy density at the quantum recollapse in such a scenario is very close to $\rcr$, unlike the previous cases discussed for 
the connection quantization. Since the quantum recollapse occurs at $V \sim 10^{24}$ (in Planck units), $\rho_4^+ \approx \rho_4^-$. 

The distinct quantum recollapses are possible only when the phantom universe undergoes quantum recollapse at not so large volumes as in Fig. 9. In that case $\rho_4^+$ and 
$\rho_4^-$ are appreciably different. Fig. 10 shows one such case for the connection quantization for equation of state $w = -10$. Instead of two types of turnarounds, as in Fig. 9, we find three distinct turnarounds. A cyclic universe composed of classical bounces and two distinct quantum recollapses which occur when $\rho_4$ becomes equal $\rho_4^+$ and $\rho_4^-$ respectively.

\subsubsection{Quantum universe initial conditions}

Let us consider the case when initial conditions are given such that the universe is not classical, and remains quantum in the entire evolution. The phantom mixed with the quantum realm brings out some 
novel features of the dynamics both in the connection and the holonomy quantization. In Fig. 11 , we depict one such representative case for the equation of state $w = -1.1$. Let us first 
discuss the curves from connection dynamics in these figures. A noticeable feature is the absence of two distinct quantum recollapses. At first this can be surprising because the volume at the recollapse is very similar to the one in the case of the simulation discussed in Fig. 10. Since the values of $\rho_4^+$ and $\rho_4^-$ do not depend on the equation of state, this apparently is a puzzling feature. One expects two distinct quantum turnarounds as in Fig. 10, yet there is only 
one type of recollapse in contrast to Fig. 10. A closer look at the features of connection dynamics reveals a very interesting feature. Two distinct quantum turnarounds are still present but one is a quantum bounce and another is a quantum recollapse! The classical turnaround does not exist. In the case of Fig. 11, we find that the quantum bounce in the connection case occurs  when $\rho_4^- \approx 0.15$. $\dot \beta$ turns out to be positive, and using eq.(\ref{dotH_conn}) so does $\dot H$ due to $D > 1$. The quantum recollapse is identified with $\rho_4^+ \approx 0.45$ (in Planck units). In this case, $\dot \beta > 0$ and $\dot H < 0$. This a very novel feature of the connection dynamics which has little in common with all the cases studied so far.

Dynamics from the holonomy quantization reveals further surprises for the case of a phantom fluid with initial conditions of the quantum universe. The curves 
appear quite similar to the connection case. In comparison to the latter, it is at first again puzzling because unlike the connection case, holonomy quantization does not permit two roots for distinct quantum turnaround energy densities. It turns out that in this case, both of the turnarounds are again `quantum'. The bounce occurs when $\rho = \rho_1$ (eq.(\ref{rho1})) as should be in the classical turnaround, albeit with a Planckian value. The recollapse occurs when $\rho = \rho_2$ (eq.(\ref{rho2})) as is for a quantum turnaround in previous cases. At the recollapse, $D$ becomes very small and so does $\rho_1$ in comparison to $\rho_2$. The latter approximates $\rcr$. $\dot \beta$ turns out to be positive and $\dot H < 0$. In contrast, at the bounce $D$ is of the order unity, and $\rho_1 \approx 0.17$. Both $\dot \beta$ and $\dot H$ turn out to be positive. It is to be noted that the bounce is not at $\rcr$ in the holonomy quantization. This effect is a result of a peculiar mixing of the phantom dynamics 
and 
the quantum universe.

To summarize, for the quantum universe initial conditions though the connection and holonomy dynamics appears very similar, there is a subtle and important 
difference between them. In the holonomy quantization, bounce occurs at $\rho = \rho_1$. Yet it is not a classical bounce. The recollapse as expected is quantum. In the connection quantization, bounce does not occur at $\rho = \rho_3$ given by eq.(\ref{rho34}) -- the expected density given similarities with the holonomy case, but instead at $\rho_4^-$. The other quantum turnaround density yields the recollpase. Finally, we find from the Fig. 11  that in the connection case the amplitude of each cycle is slightly higher, which in turn forces the solution to become ``out of phase" with the holonomy solution at late times. 

\begin{figure}[tbh!]
\includegraphics[scale=0.48]{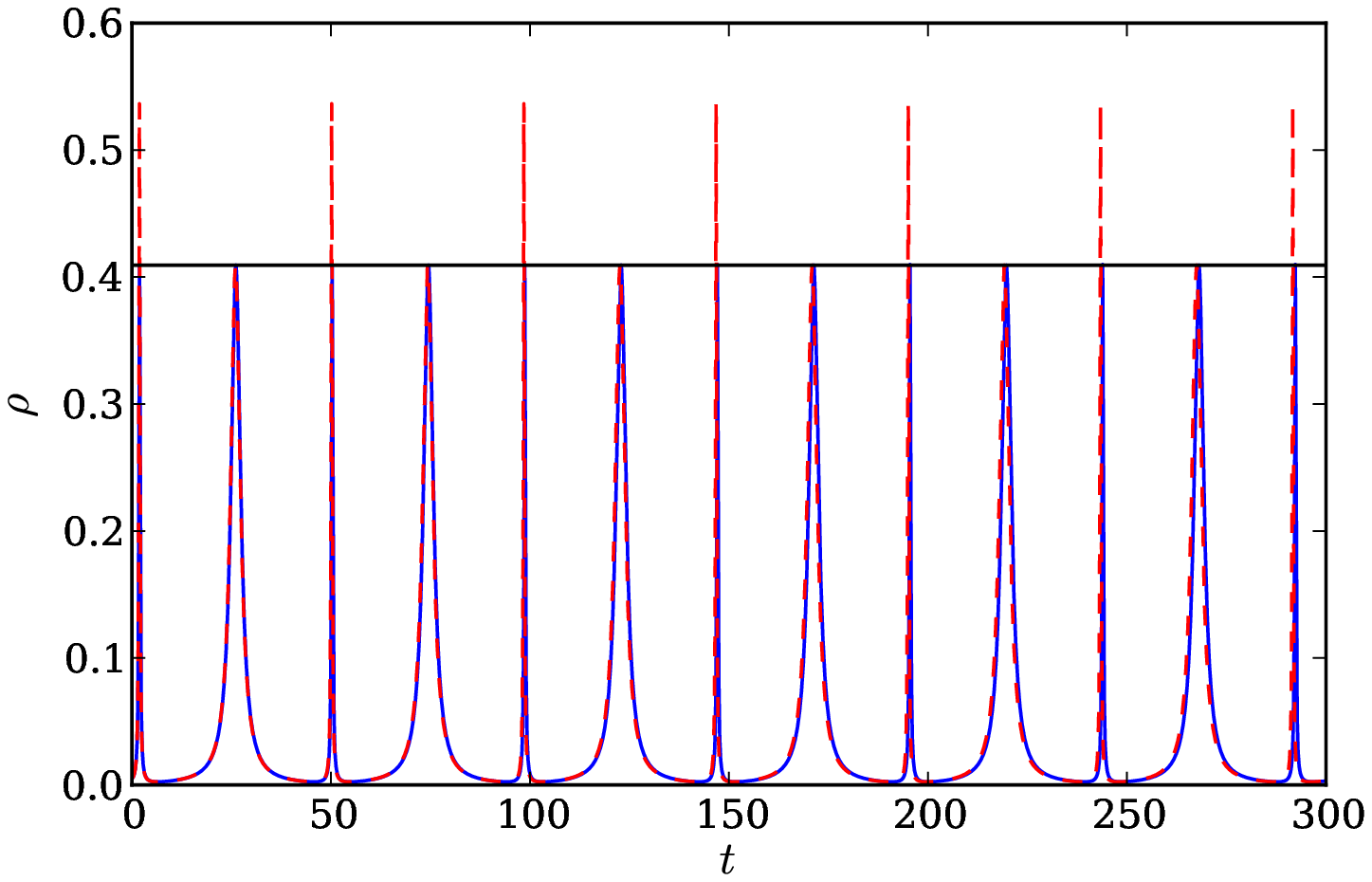}
\includegraphics[scale=0.48]{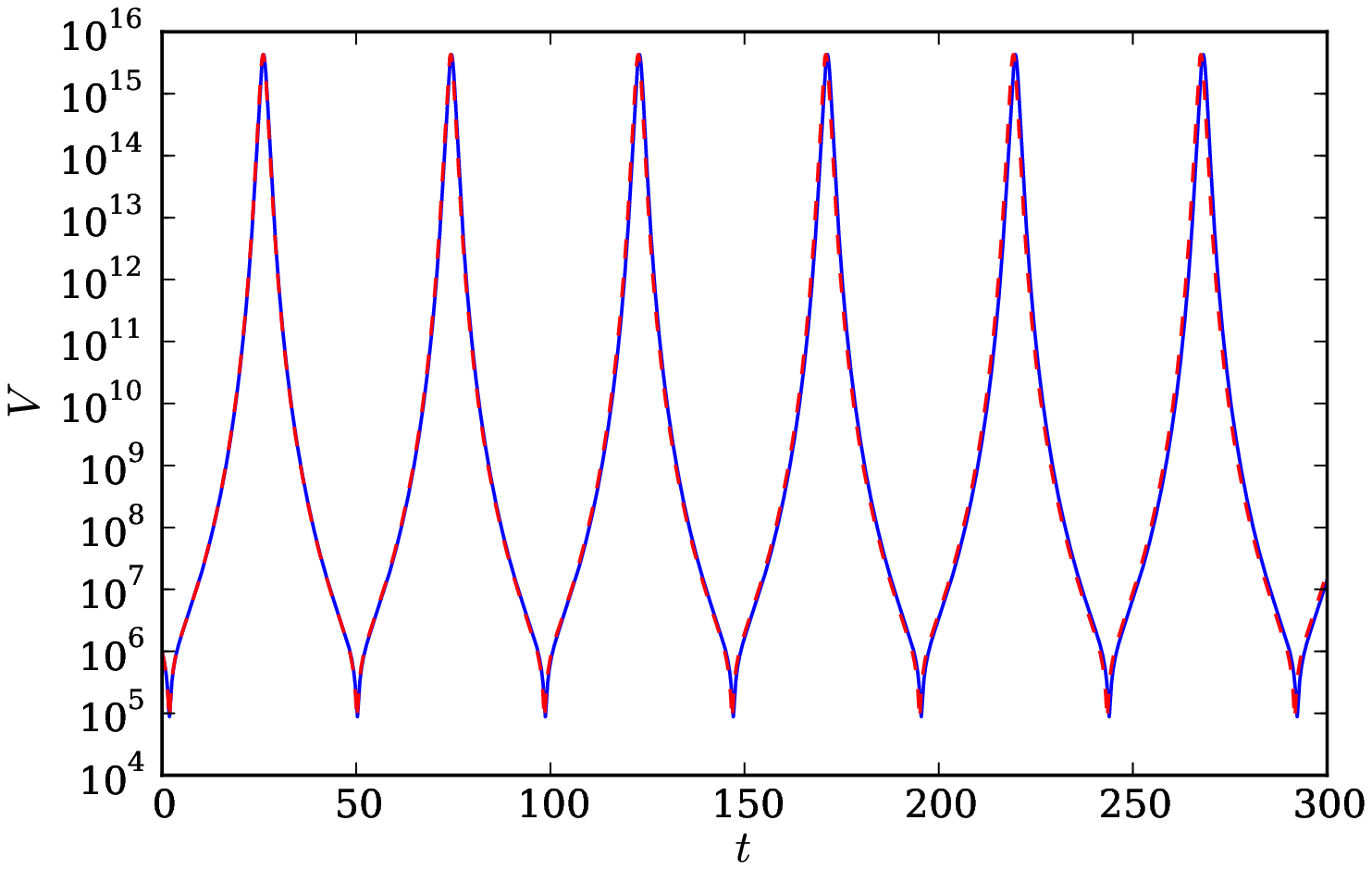}
\caption{For a two fluid scenario with $w_1 = 1.0$ and $w_2 = -1.25$, behavior of energy density and volume are shown. Initial conditions are  $V_0=10^6$, $\rho_0 = 0.0016$, such that $\rho_{0,\rm tot}= 0.0032$. The horizontal black line represents $\rho_{\rm max}$. The solid curves correspond to holonomy quantization and dashed curves correspond to connection quantization.  }
\end{figure}

\subsection{Two Fluid Universes}

Depending on the equation of states of the individual fluids, rich dynamics originates in the two fluid scenario when the phantom fluid is present. In the following we present a solution which captures a representative case which results in an interesting phenomenology with two classical singularities: a two fluid scenario composed of a stiff matter and a phantom fluid. The stiff matter dominates only near the classical big bang singularity, and the phantom dominates elsewhere. As a result, in the classical theory, there are two singularities. A big bang in the past and a big rip in the future. Quantum geometric effects 
in the holonomy as well as in the connection quantization resolve both of these singularities, resulting in a cyclic universe. Instead of the three distinct turnarounds in the connection dynamics as discussed for Fig. 10, we obtain only two turnarounds. Both are quantum in nature. One is a quantum bounce which resolves the classical big bang singularity, and another is a quantum recollapse which resolves the classical big rip singularity. In the holonomy case, a cyclic universe again originates with two quantum turnarounds -- one bounce and one recollapse.

Fig. 12, demonstrates the case discussed above. Initial conditions are chosen for the classical contracting universe. The equations of state correspond to 
$w_1=1.0$ and $w_2=-1.25$. In the contracting phase, stiff matter 
quickly dominates the dynamics and the energy density grows very high very quickly as the volume decreases. Note that $\rho_1 \propto a^{-6}$ and $\rho_2 \propto a^{3/4}$. In the holonomy quantization the energy density grows to $\rcr$ and then bounces. This is in stark contrast to the solution we considered in the single fluid case in Fig. 11, where the energy density was at a value of 0.17 when the quantum structure and phantom energy forced turnaround. In the present case, the bounce is purely kinetic dominated (or due to the massless scalar) and the phantom energy plays little role. There is then a period of inflation and energy density grows as the volume increases due to the phantom energy dominating the overall energy density. The energy density grows to $\rcr$ and then there is another bounce from the quantum structure back into a contracting phase. The cycle then repeats and the minimal volume obtained is approximately the same in each cycle.

In the connection quantization of Fig. 12, dynamics results in bounces and recollapses (both quantum). The bounce occurs when energy density equals $\rho_4^+$ which takes the values approximately equal to 0.52 in Planck units. The recollapse occurs at very large volume and in that case $\rho_4^+ \approx \rho_4^-$. Since $D$ is negligible in this case, the quantum recollapse occurs at approximately $\rcr$. The similarity with the case discussed in Fig. 11 should be noted. There are only two turnarounds, not three. Both the turnarounds are quantum in nature and one of the quantum turnarounds corresponds to the bounce and other to the recollapse.

\section{Probing the Super-Planckian Regime}

In the previous sections, we have discussed the dynamics in the holonomy and connection quantization assuming validity of effective dynamics for a variety of cases. In the connection quantization, many simulations result in a very large value of the energy density at quantum recollapse. Though not common in the holonomy quantization, universes which retain their quantum character, and those which have phantom component give hints that extreme energy densities can also occur for 
holonomy quantization. In particular, in such cases energy density can take value larger than $\rcr$ which is sometimes considered to be phenomenologically 
a bound on the energy density in LQC. Though strictly speaking this is true only for the spatially flat model. In this section, we consider this extreme quantum regime for the connection and the holonomy quantization in more detail. We also discuss the properties of the expansion scalar, which unlike the energy density, is universally bounded in the holonomy quantization. We start with a discussion of how extreme energy densities are phenomenologically possible in 
connection quantization in single fluid and two fluid scenarios, and remark on breaching of $\rcr$ by energy density in holonomy case.

\subsection{Single Fluid Universes}
An example of the dynamics probing the super-Planckian regime is shown in Fig. 13 in which the volume probes very close to zero in the connection dynamics and the energy density probes super-Planckian values. The equation of state corresponds to $w=-1/3$ which in particular has an interesting effect on the connection quantization.

\begin{figure}[tbh!]
\includegraphics[scale=0.48]{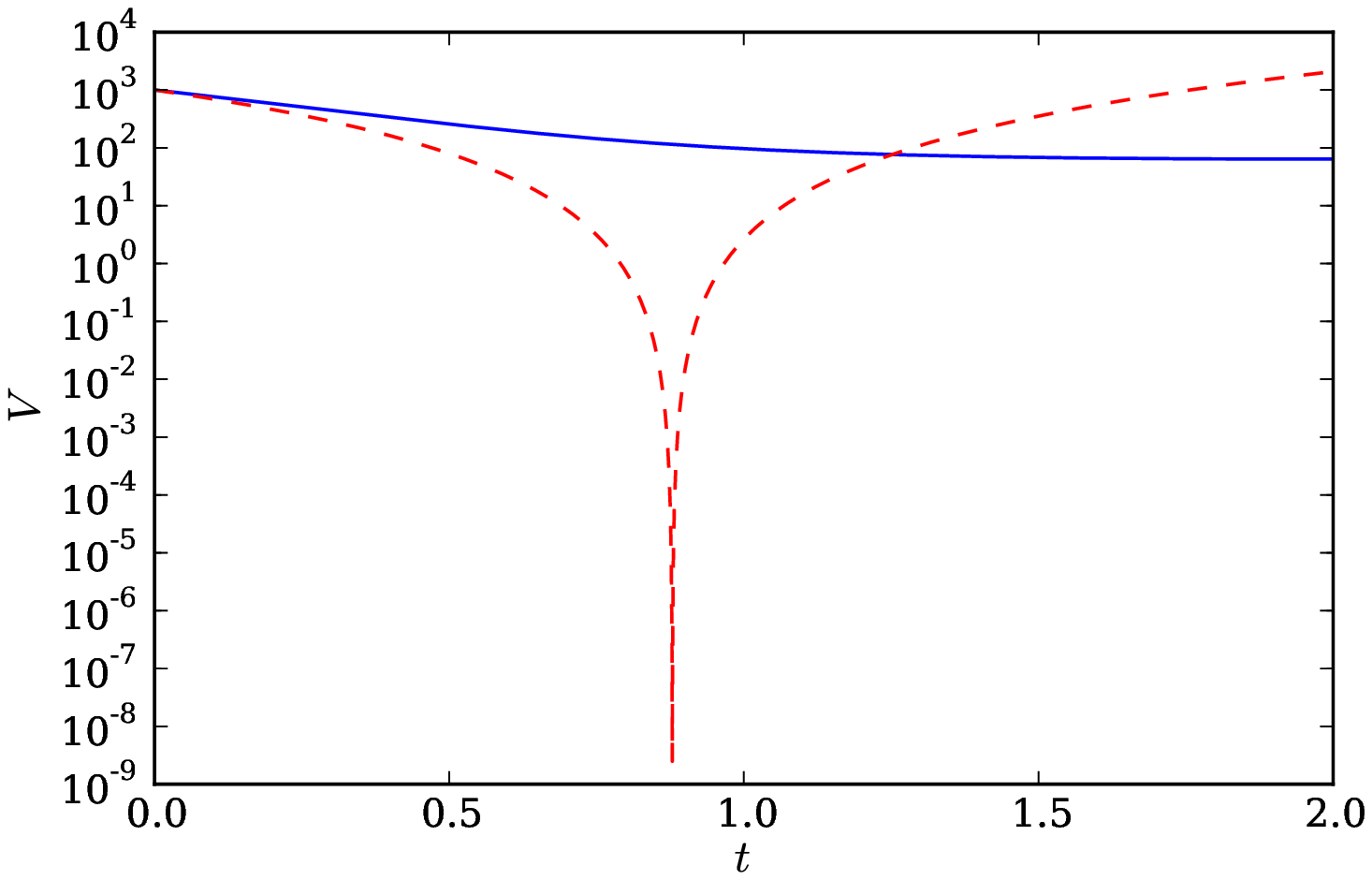}
\includegraphics[width=7.5cm, height=5cm]{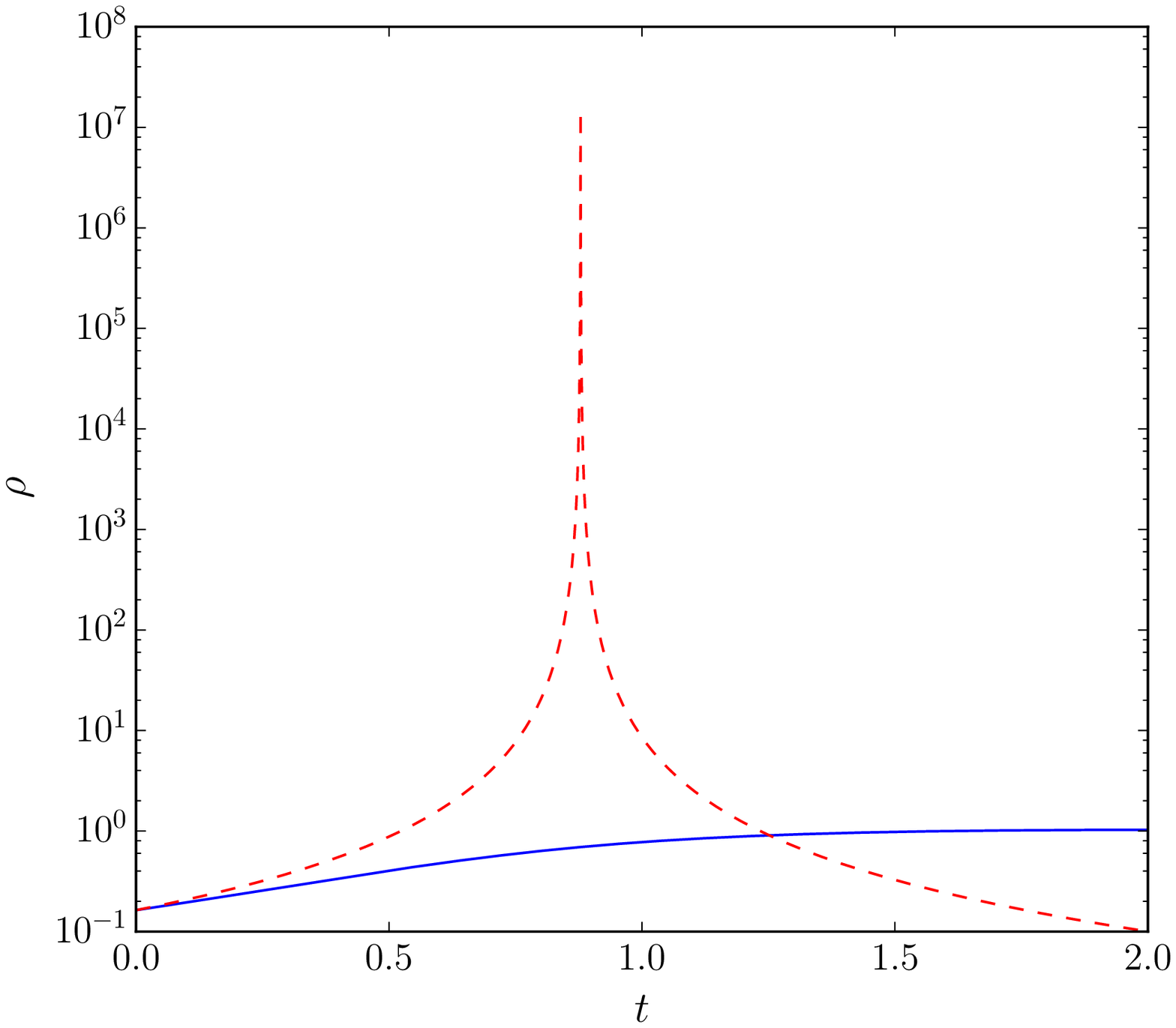}
\caption{Evolution of volume and energy density versus time for the case of $w=-1/3$ is shown for the holonomy (solid curve) and connection (dashed curve) quantization. Initial conditions in Planck units for this run were as follows: $V_0 = 1000$ and $\rho_0 = 0.16335$. The volume at the bounce in the case of holonomy quantization is $V_b \approx 64 V_{\rm{Pl}}$.}
\end{figure}

We start with a contracting universe with highly quantum initial conditions. We select a very high initial energy density and a correspondingly low initial volume. This, combined with the selection of the equation of state forces the connection quantization toward super-Planckian energy densities and a low volume very quickly. We see that the volume reaches $10^{-9}$ (in Planck units), where the effective spacetime description is expected to break down. Singularity is still resolved because the underlying quantum difference equation is non-singular. Due to the 
bounce at such a small volume, the energy density at the bounce is around $10^7 \rcr$! Such high densities are never seen in the holonomy quantization where the maximum energy density is usually on the order of $\rho_{\rm max}$. After the bounce,  the connection solution undergoes accelerated expansion and a turnaround is no longer possible as the energy density of the universe decreases as the volume increases. In contrast, during this period the holonomy dynamics does not produce anything extraordinary. In the subsequent time evolution than the one shown in Fig. 13, holonomy quantization results in a bounce at approximately 
the energy density equal to $\rcr$.

It should be noted that the error in satisfying the Hamiltonian constraint in the connection quantization grows by three orders of magnitude, i.e. from $10^{-9}$ to $10^{-6}$ immediately following the bounce. This is due to the extraordinary rate at which $V$ and its derivatives are changing in time. It should also be noted that Fig. 13 shows one of the extreme examples found in our study. With different initial conditions, the effect of reaching almost vanishing volume is not nearly as dramatic. However, energy densities with super-Planckian values are not difficult to obtain.

\subsection{Two Fluid Universes}
 
As in the case of the single fluid scenario, it is not difficult to approach very close to the classical singularity in the two fluid scenario in the 
connection quantization. In this regime, effective spacetime description can be suspected to reach its limit of validity, and the inverse volume modifications are expected to play an important role in this very small regime near the classical singularity. Following the single fluid case discussed above, it is straightforward to obtain cases where the universe in the connection quantization repeatedly achieves super-Planckian energy densities if the effective spacetime is assumed to be valid. Instead of the eternal accelerated expansion after the bounce in Fig. 13 in connection quantization, we can achieve a quantum recollapse by adding a phantom fluid. As a result the universe undergoes cyclic behavior. 
 
In Fig. 14 we present a solution with the same initial conditions as the previous single fluid case but with a turnaround due to a phantom energy component. The equations of state are $w_1=-1/3$ and $w_2=-1.25$, corresponding to energy densities $\rho_1 \propto a^{-2}$ and $\rho_2 \propto a^{0.75}$. Again, because the derivative of the volume very close to the bounce is quite high, and the derivative changes very rapidly at the bounce, these simulations produce numerical errors in the Hamiltonian constraint that oscillate from $\sim 10^{-10}$ to $\sim 10^{-6}$ up to $t=50$. These errors account for the observed numerical differences in the value of $V$ at each bounce.        
       
\begin{figure}[tbh!]
\centering
\includegraphics[scale=0.47]{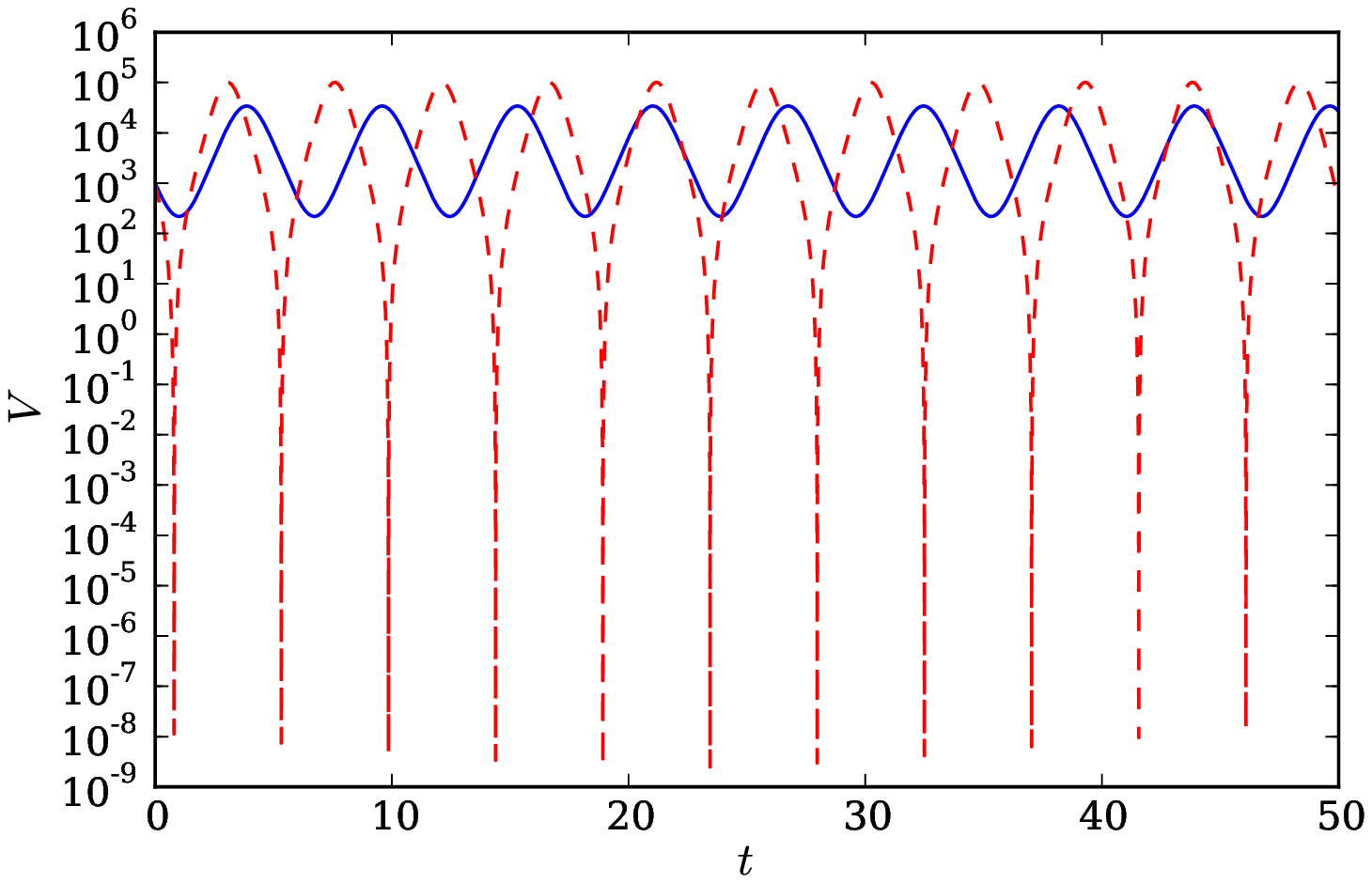}
\includegraphics[scale=0.47]{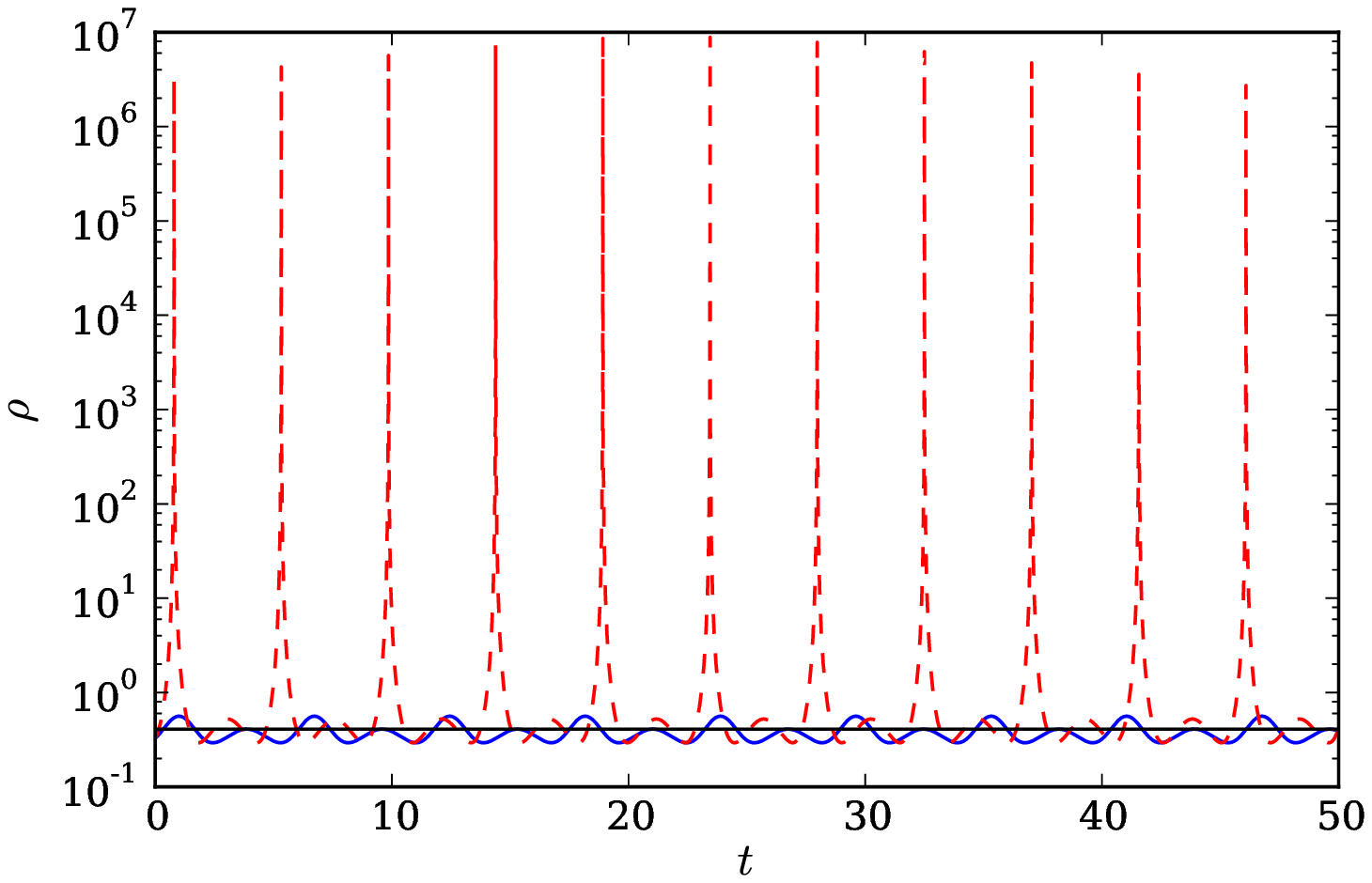}
\caption{$V$ vs. $t$ for the same initial conditions as Fig. 13, for two fluids corresponding to $w_1 = -1/3$ and $w_2 = -1.25$. }
\end{figure}   

From Fig. 14 we also see that in the connection quantization the energy density spikes at the bounce, followed by a rapid period of accelerated expansion where the energy density falls dramatically. In the following evolution, two fluids compete with each other with the phantom fluid winning at larger volumes causing an increase in energy density as $V$ increases. The energy density again shoots up till quantum geometric effects take over and cause a quantum recollapse. The recollapse occurs at energy density approximately equal to $\rcr$. The universe enters a contraction in the further evolution and the cycle repeats.

Surprisingly, another important aspect of these simulations is the holonomy quantization, in which $\rho$ oscillates to a value slightly greater than $\rcr$ when the universe bounces. This effect is not seen for the majority of initial conditions explored in this study, however, as we will see shortly, it is not forbidden. In contrast to the connection quantization, dynamics in the holonomy quantization results in smooth cyclic behavior with energy density oscillating close to $\rcr$.  

It should be noted that in the above analysis for single fluid and two fluid scenarios we have ignored the inverse 
volume modifications. In presence of the latter modifications, the energy density and the expansion scalar in the connection quantization turns out to be bounded. However, the theoretical bound is super-Planckian and is estimated to be approximately $26.8 ~ \rcr$ \cite{ck-closed2}. Above simulations in our phenomenological analysis provide an evidence of such super-Planckian values being reached in the connection quantization. Since our analysis ignores inverse volume modifications, the energy densities at bounces turn out to be much higher than the above estimate. We expect that an analysis where such modifications are included, would alleviate such energy densities to lower but still super-Planckian values.

\subsection{$\boldsymbol{\rho > \rcr}$ in Holonomy Quantization}

In the two fluid scenario discussed above we found that holonomy quantization results in energy density at the bounce which is greater than $\rcr$. This is the first example of a dynamical solution resulting in such energy densities at the bounce for the holonomy quantization. Though at first surprising, this breach is easy to understand. We first note that unlike $\rcr$ in the spatially flat case, the energy density at the quantum turnaround in the holonomy quantization is not universally bounded. Energy densities in holonomy quantization of $k=1$ model greater than $\rcr$ can occur due to large value of $D$ which results in large enough $\rho_1$ (\ref{rho1}). 

In the holonomy quantization though the energy density is not universally bounded, the expansion scalar is. In fact, in the  absence of inverse volume modifications there is no bound on energy density at all in the holonomy quantization. This is in contrast to the expansion scalar, whose bound is given by $|\theta_{\rm max}| = 3/2 \gamma\lambda$. In the spatially flat case, the bound on the expansion scalar, which is same as the above value, implies a bound on $\rho$ given by $\rcr$. The situation is not the same in the spatially closed model. In the absence of inverse volume modifications $\rho$ can extend beyond $\rcr$ in the holonomy quantization, even while the bound on the expansion scalar is still satisfied. This result suggests that expansion scalar is more important than the energy density if we wish to understand 
the Planck scale phenomenology in LQC in terms of a bounded variable. Note that if the inverse volume modifications are considered then the energy density in holonomy quantization gets bounded, with a theoretically estimated bound approximately $18.63 \rcr$ \cite{ck-closed2}. As we will discuss now, the simulations in Fig. 15 clearly show that phenomenologically cases where $\rho$ is greater than $\rcr$ are possible.

Shown in Fig. 15 are the energy density and expansion scalar as a functions of time for a universe with highly quantum initial conditions 
with a massless scalar field ($w=1$). We can see that in the holonomy quantization $\rho$ spends more time above $\rcr$ than below it, and 
that a solution exists even if $\rho_0 > \rho_{\rm max}$.  Relative errors in the solution remain very small in the plotted region, 
remaining around $10^{-8}$. This universe is essentially stuck in highly quantum state, oscillating from maximum to minimum volume on very 
small time scales. Interestingly, in many other instances where the initial conditions are not so extreme, $\rho$ consistently probes energy 
densities greater than $\rho_{\rm max}$. We observe this across the board as $\rho_0$ is varied from 0.05 to 0.3, and $V_0$ is kept at 1000 
for a variety of values of equation of state parameter.


\begin{figure}[tbh!]
\centering
\includegraphics[scale=0.47]{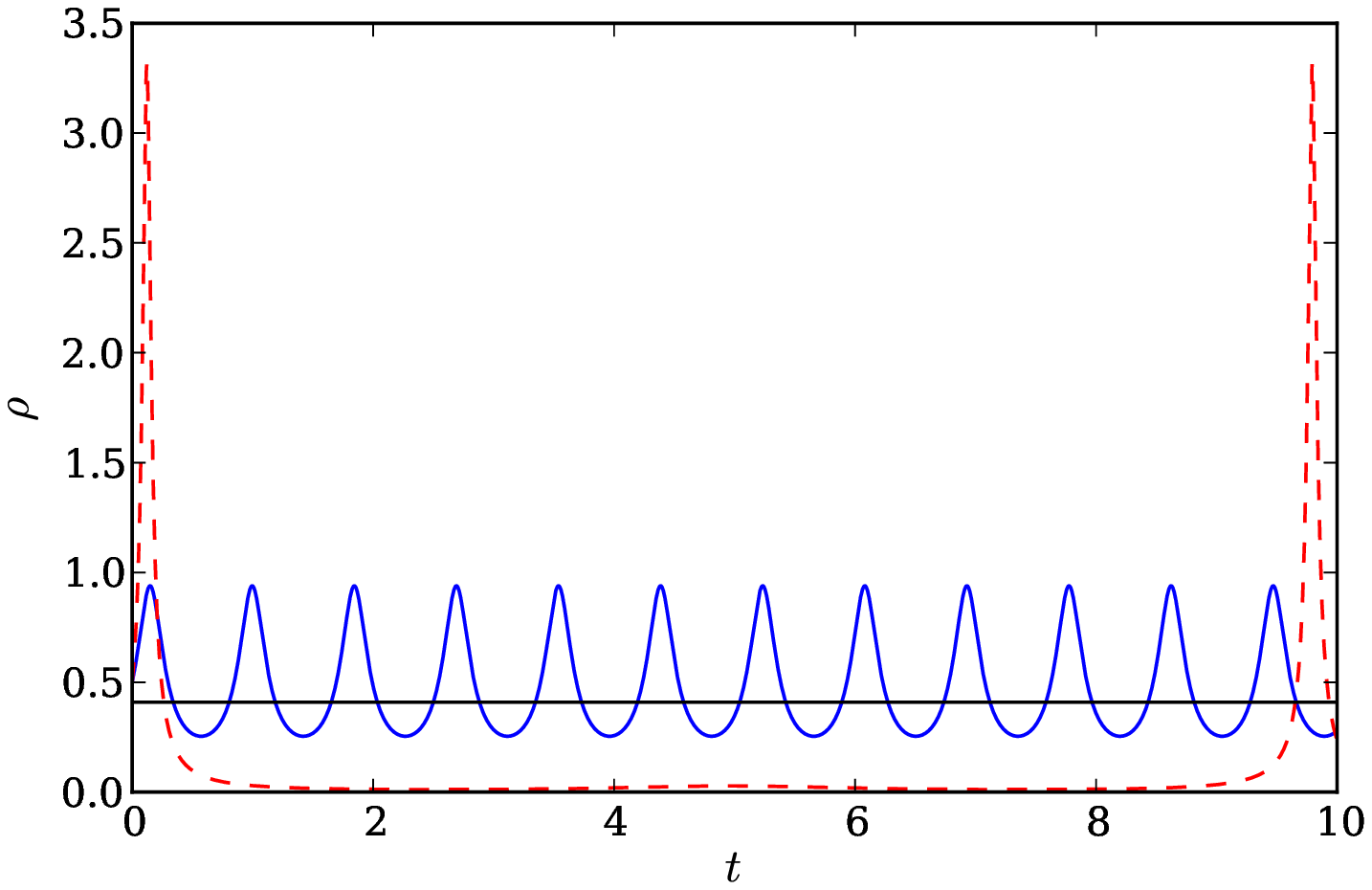}
\includegraphics[scale=0.47]{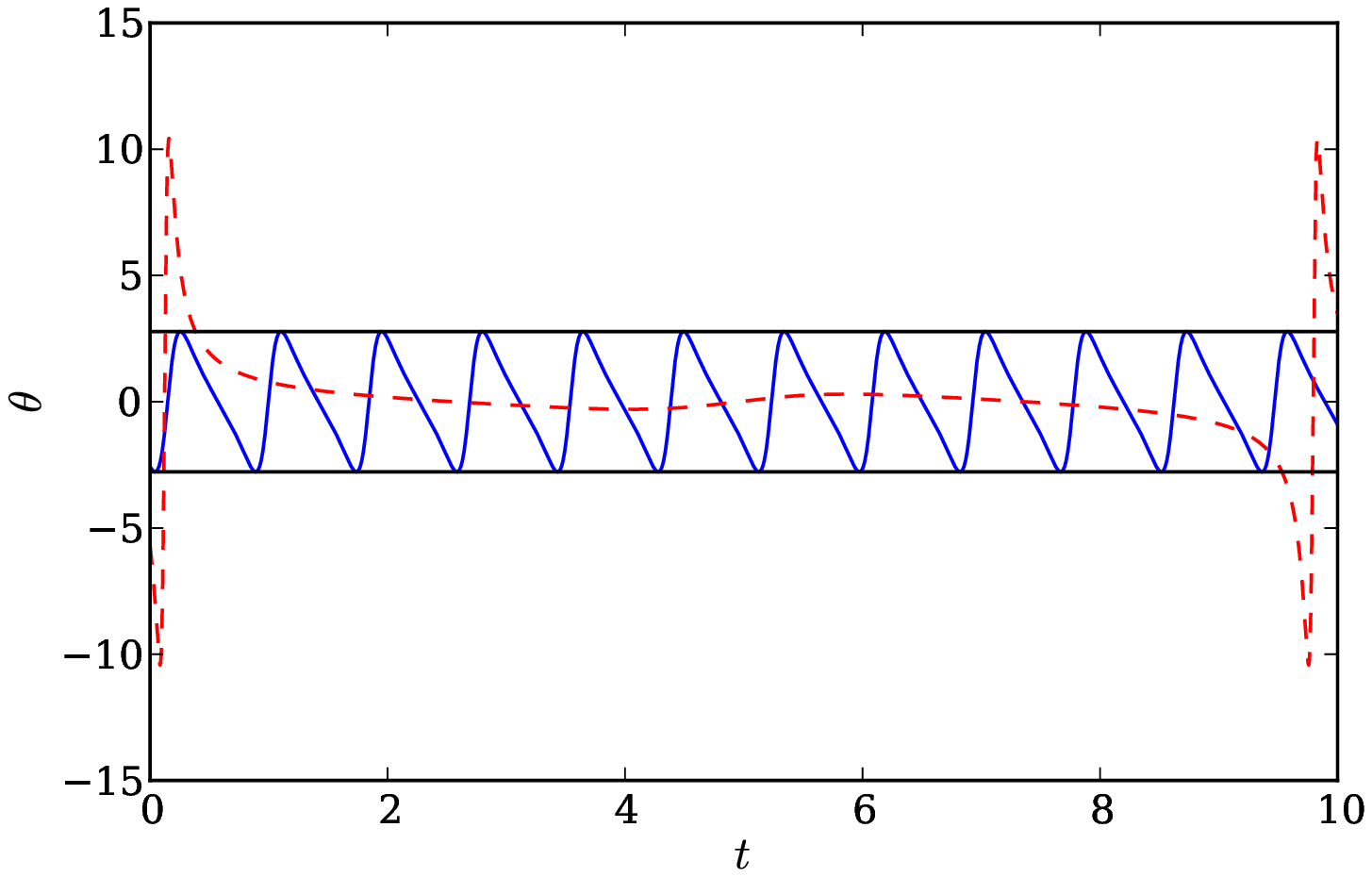}
\caption{Plot of energy density  and expansion scalar versus time for a massless scalar field ($w=1$) with initial conditions $\rho_0=0.5 > \rho_{\rm max}$ and $V_0 = 100$. The black horizontal line represents $\rcr$ in the left figure and the maximum allowed value of expansion scalar in the holonomy quantization in the right figure. The holonomy solution is represented by the solid line while the connection solution is represented by the dashed lines.}
\end{figure}

In contrast to the dynamics in the holonomy quantization, the connection quantization results in very large value of energy density at quantum turnaround. The expansion scalar in connection quantization also takes a very high value at these turning points. Note that the expansion scalar in the connection quantization takes very large values even when one includes inverse volume modifications. In particular, in the presence of inverse volume modifications the bound on expansion scalar is approximately $|\theta| = 23.6$ in Planck units. In contrast, the expansion scalar is bounded in holonomy quantization by $|\theta| \approx 2.78$ in Planck units. Thus, even though the inverse volume modifications bound the expansion scalar in connection quantization, the corresponding value can be much larger than the one for the holonomy quantization. Our results in Fig. 15 provide a phenomenological evidence of this large disparity between the values of expansion scalar in two quantization schemes.

As explained at the end of Sec. II, the energy density at the quantum turnaround in the connection quantization is generally greater than in the holonomy quantization if the turnarounds occur at the same volume. As we discussed earlier, recall that in all these cases, there is only one quantum bounce in the connection case. The energy density at this quantum turnaround corresponds to $\rho_4^+$ which is always greater than $\rho_1$. In the above simulation, the bounce volume in the connection case is approximately $V = 39 V_{\rm{Pl}}$ which is smaller than the bounce volume in the holonomy case which is approximately $V = 73  V_{\rm{Pl}}$. This acts like an additional cause to drive the bounce volume in connection quantization to a higher value than the holonomy quantization. The same is true for the case of the simulation in Fig. 13. 


\section{Conclusions}
In quantization of spacetimes in given framework, a pertinent issue is how different the qualitative predictions are in different consistent quantizations. This issue, which is important for phenomenological considerations, so far has received less than the deserved attention in LQC. 
Only recently have these issues started to be rigorously explored in Bianchi model and Kantowski-Sachs spacetimes \cite{bgps-spatial,pswe,cs-schw,ks-constant,ps16a}. In the isotropic case, 
for the spatially flat spacetime no two consistent quantizations exist \cite{cs-unique,cs-geom}. However, this is not the case for the spatially closed model, where apart from the earlier quantization based on the closed holonomies \cite{apsv,closed-warsaw}, a new consistent quantization based on the connection operator using open holonomies was proposed recently by Corichi and Karami \cite{ck-closed1}. The latter quantization is based on the techniques used to quantize Bianchi-II  and Bianchi-IX spacetimes in LQC \cite{awe-bianchi2,we-bianchi9,ck-b9}. To study the qualitative differences between the holonomy and the connection operator quantization methods, the $k=1$ FLRW model provides a very useful setting. Some of the qualitative differences were known from earlier works. These include 
existence of two distinct quantum bounces for the connection quantization instead of one for the holonomy quantization, explicitly shown for the case of a massless scalar field \cite{ck-closed1}. The difference in the properties of the expansion scalar in the two quantizations was also studied in Ref. \cite{bgps-spatial}. However, a detailed phenomenological analysis of the qualitative differences for different types of matter content between the two methods was so far missing. One more important gap was the lack of study of universes with quantum initial conditions and the way such a choice affects the predictions. Further, so far only a little was known about the super-Planckian regime in the connection quantization approach, and super-Planckian regime in the holonomy quantization, though analytically not ruled out, was never investigated till now.

In this manuscript, our goal was to fill these gaps in the understanding of the physics of $k=1$ FLRW models in LQC. We considered single fluid and two fluid scenarios for matter with various equations of states including phantom matter. Initial conditions corresponding to evolution starting in classical as well as quantum realms were considered. A large number of simulations were performed of which representative cases have been discussed in this paper. Primary objectives of this manuscript can be broadly classified as: to understand the robustness of the two distinct quantum bounces in the connection quantization approach, to gain insights on the number -- whether two or three, and the nature of turnarounds -- whether classical or quantum in different phenomenological settings, and to understand the super-Planckian regime both in the connection and the holonomy quantization methods. Note that we focused on the modifications to the effective Hamiltonian coming from expressing field strength or the 
connection operator in terms of holonomies. Modifications originating from the inverse volume terms, which generally do not dictate dynamical evolution, were ignored. Inclusion of the inverse volume modifications might change certain quantitative details about the results in the super-Planckian regime. As discussed in the manuscript, these modifications affect the Hamiltonian constraints in the two quantization prescriptions in the same way. If the bounce occurs at volumes much greater than the Planck volume, the inverse volume modifications have a negligible effect. Further, even though the energy density in the holonomy case, and the energy density and the expansion scalar in the connection case are estimated to be bounded \cite{ck-closed2}, these bounds are super-Planckian and do not affect our results qualitatively. Our numerical simulations provide an evidence that such super-Planckian bounds can be achieved in phenomenology. 

Let us summarize the main results from our investigations. We first show that for matter satisfying strong energy condition, the connection quantization results in a non-singular cyclic scenario with two distinct quantum bounces and a classical recollapse. The differences between the two distinct quantum bounces and the nature of evolution around them depends on the equation of state as well as the initial conditions. These results generalize earlier results in the connection quantization for the phenomenological studies of massless scalar field \cite{ck-closed1}, and agree with the general conclusions of the latter work.  In contrast, in the holonomy quantization matter satisfying the strong energy condition, a cyclic scenario with a single 
distinct quantum bounce and a classical recollapse occurs. Our next result concerns fluids which violate the strong energy condition but still obey the weak energy condition. In the case of a single fluid scenario which violates the strong energy condition, there is only 
a single distinct quantum bounce in the holonomy and the connection quantization methods. The reason is tied to the eternal accelerated expansion of the universe which commences after the first bounce due to which the second bounce does not occur in the evolution. The case of two fluid scenarios is richer. If one of the fluid satisfies the strong energy condition, then depending on the initial energy densities cycles can repeat and distinct quantum bounces reappear in the connection quantization. In the holonomy quantization, such cycles, as expected, only yield a single distinct bounce. Thus, two distinct quantum bounces is a robust phenomena for the single fluid and the two fluid scenarios satisfying the strong energy condition in the connection quantization. But it is absent if the fluids violate strong energy condition (unless they have a huge violation of the weak energy condition as becomes evident for the phantom fluids).

Presence of phantom fluids brings out some novel features of loop quantum dynamics, both in the holonomy and connection quantization 
schemes. Unlike in GR, there is no big rip but a ``big recollapse.'' Unlike for fluids which obey the weak energy condition, the bounce 
need not be quantum. Let us first discuss the cases when initial conditions correspond to classical universes. For such universes, sourced 
with a single fluid, which 
become classical in the evolution, a bounce is purely classical. It originates from a violation of the weak energy condition by the phantom fluid. Quantum geometry instead leads to recollapses which prevent the classical big rip singularity to occur. The universe is cyclic both in the connection operator and the holonomy quantizations. Generally, the quantum recollapse occurs at very large volume because of which the energy densities of the two distinct quantum recollapses in the connection quantization become indistinguishable. As a result, there is 
essentially only a single distinct quantum turnaround in the connection quantization. Distinct quantum bounces only appear if one considers an equation of state which is much less than negative unity. In such a case, strong phantom equation of state drives the universe towards the big rip very quickly and the quantum geometry effects start playing role at much smaller volumes than when phantom equation of state is weak. Thus, for single fluid universes which become classical during the evolution, the connection quantization yields strikingly different results for matter which satisfies the strong energy condition and which violates the weak energy condition. Instead of three distinct turnarounds, a classical recollapse and two quantum bounces, in the former, one obtains only two distinct turnarounds, a classical bounce and a quantum recollapse, in the latter unless the matter has very strong phantom nature, i.e. $w \ll -1$. The situation in the holonomy quantization is the same as the connection quantization 
in 
the phantom case for universes with classical initial conditions and not having  an extreme phantom equation of state. This is the only case where there are no phenomenological differences between the holonomy and connection quantization approaches. The two fluid scenario results in a rich mix of dynamical features that depend on initial conditions. If the equation of state 
of one of the fluids is phantom and another fluid satisfies the strong energy condition, then classically there is a big bang as well as a big rip singularity. 
Both the singularities are avoided in LQC in the holonomy as well as the connection quantization approaches. In such a scenario, two distinct quantum bounces reappear in the connection quantization.

We found some interesting results for the universes which remain quantum in the case of the phantom fluid. Instead of three distinct turnarounds we find only two turnarounds, both quantum. There is no classical turnaround in this case.  The novelty of this phenomena is added by the observation that unlike any other cyclic model in a phantom universe, the bounce and the recollapse originate due to quantum geometry. And unlike any other case studied so far in connection dynamics and the holonomy dynamics, the quantum turnarounds are neither bounces nor recollapses, but both. Though the holonomy and connection quantization yield a very similar picture of a single quantum bounce and single quantum recollapse there are some striking differences. In the case of the connection quantization, the bounce and recollapse are given by the two 
roots of the quantum turnaround. Whereas in the holonomy case, the quantum bounce and the quantum recollapse originate from the single and the only root for the quantum turnaround. The latter case gives valuable insights in the holonomy quantization on the values of energy density getting larger than the maximum value in spatially flat model in LQC. 

Dynamics of the connection quantization case shows that it is easy to reach situations of the super-Planckian regime. Sometimes very extreme values can be reached and universe in such a case probes extremely closely to the classical big bang singularity. In the connection quantization, not only the energy density but the expansion scalar can also attain very high super-Planckian values. Generally, dynamics lead to bounce densities which are larger than maximum values of energy density in the spatially flat isotropic model in LQC. The volume at which bounce occurs becomes much smaller than the Planck volume. In this regime, inverse volume modifications to the effective Hamiltonian dynamics are expected to play an important role. Surprisingly, initial conditions where energy density at the bounce can be larger than the maximum energy density in the spatially flat FLRW model in LQC regime also exists for the holonomy 
case. Though not ruled out analytically, previous studies of the closed FLRW model in LQC 
have not found this regime phenomenologically. The expansion scalar in this case turns out to be dynamically bounded by the universal value. This result suggests that for the spatially curved models in the 
holonomy quantization, the more canonical measure of the Planck regime in LQC should be the value of the expansion scalar rather than the energy density.  As we discuss in the manuscript, bounds on energy density and expansion scalar in the connection approach, and the energy density in the holonomy approach are obtained if we include the inverse volume modifications. These bounds though are super-Planckian, are expected to alleviate extreme values we found in the simulations.  Further, the validity of the effective spacetime regime for such small values of volume can be questioned. To fully understand the effects of inverse volume modifications, it is important to perform numerical simulations with quantum states and carefully analyze and compare the effective dynamics .

Finally, let us comment on how these results can prove useful to narrow down the quantization ambiguities in the spatially closed model in LQC. Working within the caveat of assuming the validity of effective spacetime description and ignoring the inverse volume effects (which turn out to be negligible when bounce densities are larger than Planck volume), we find that in the connection quantization there exist many initial conditions for which energy density would attain super-Planckian values. Though super-Planckian values for holonomy case are possible, such initial conditions only occur in highly quantum cases and are also rare. Even if we include estimates from including inverse volume modifications, we expect our results to be not affected qualitatively because the bounds on energy density and expansion scalar in the connection case are super-Planckian. These results suggest that there is potentially a significant  quantitative difference in the physics of Planck regime in the connection and the holonomy 
quantizations. Future phenomenological investigations, such as those probing cosmological perturbation spectrum in models based on these spacetimes,  can help in showing whether this difference can constrain one of these quantization prescriptions.  However, it is important to understand the validity of the effective spacetime description in this framework along with the inclusion of the inverse volume effects in loop quantizations of $k=1$ model. In a phenomenological study based on the latter, interpretation and role of the effective equation of state becomes non-trivial due to inverse volume effects \cite{ps05}. It will   be interesting to investigate the detailed phenomenology in effective spacetime with such effects to gain in further insights on trans-Planckian effects for different fluids. Further, detailed numerical simulations using quantum states, on the lines of recent works\cite{numlsu-2,numlsu-3}, will prove useful to shed further insights on these issues. Numerical simulations with physical 
states in these models will be able to provide important clues on the validity of effective description, its regime of validity and the role of inverse volume modifications. In conclusion, though one may be tempted to to state that the holonomy quantization perhaps yields a more viable physical description than the connection quantization where super-Planckian energy densities are often possible, the final answer depends on numerical simulations where evolution of quantum states is studied using quantum Hamiltonian constraint for the holonomy and connection quantizations.

\section*{Acknowledgments}
We are grateful to Alejandro Corichi for valuable discussions and comments on the manuscript, and thank S K Soni for comments. We thank an anonymous referee for helpful comments which improved this manuscript. 
This work is supported by NSF grants PHY-1404240 and PHY-1454832. J.D. thanks the REU program in Physics and Astronomy at LSU during
which a large part of this work was completed.






\begin{thebibliography}{10}
\bibitem{as-status}
A.~Ashtekar and P.~Singh,
\newblock Class. Quant. Grav. {\bf 28}, 213001 (2011), arXiv:1108.0893.

\bibitem{cs-unique}
A.~Corichi and P.~Singh,
\newblock Phys. Rev. {\bf D78}, 024034 (2008), arXiv:0805.0136.

\bibitem{cs-geom}
A.~Corichi and P.~Singh,
\newblock Phys. Rev. {\bf D80}, 044024 (2009), arXiv:0905.4949.

\bibitem{aps1}
A.~Ashtekar, T.~Pawlowski, and P.~Singh,
\newblock Phys. Rev. Lett. {\bf 96}, 141301 (2006), arXiv:gr-qc/0602086.

\bibitem{aps2}
A.~Ashtekar, T.~Pawlowski, and P.~Singh,
\newblock Phys. Rev. {\bf D73}, 124038 (2006), arXiv:gr-qc/0604013.

\bibitem{aps3}
A.~Ashtekar, T.~Pawlowski, and P.~Singh,
\newblock Phys. Rev. {\bf D74}, 084003 (2006), arXiv:gr-qc/0607039.

\bibitem{slqc}
A.~Ashtekar, A.~Corichi, and P.~Singh,
\newblock Phys. Rev. {\bf D77}, 024046 (2008), arXiv:0710.3565.

\bibitem{consistent-lqc}
D.~A. Craig and P.~Singh,
\newblock Class. Quant. Grav. {\bf 30}, 205008 (2013), arXiv:1306.6142.

\bibitem{consistent-vertex}
D.~Craig and P.~Singh,
\newblock {The Vertex Expansion in the Consistent Histories Formulation of Spin
  Foam Loop Quantum Cosmology},
\newblock in {\em {14th Marcel Grossmann Meeting on Recent Developments in
  Theoretical and Experimental General Relativity, Astrophysics, and
  Relativistic Field Theories (MG14) Rome, Italy, July 12-18, 2015}}, 2016,
  arXiv:1603.09671.

\bibitem{ps12}
P.~Singh,
\newblock Class. Quant. Grav. {\bf 29}, 244002 (2012), arXiv:1208.5456.

\bibitem{khanna-review}
D.~Brizuela, D.~Cartin, and G.~Khanna,
\newblock SIGMA {\bf 8}, 001 (2012), arXiv:1110.0646.

\bibitem{apsv}
A.~Ashtekar, T.~Pawlowski, P.~Singh, and K.~Vandersloot,
\newblock Phys. Rev. {\bf D75}, 024035 (2007), arXiv:gr-qc/0612104.

\bibitem{closed-warsaw}
L.~Szulc, W.~Kaminski, and J.~Lewandowski,
\newblock Class. Quant. Grav. {\bf 24}, 2621 (2007), arXiv:gr-qc/0612101.

\bibitem{bp-lambda}
E.~Bentivegna and T.~Pawlowski,
\newblock Phys. Rev. {\bf D77}, 124025 (2008), arXiv:0803.4446.

\bibitem{kv-open}
K.~Vandersloot,
\newblock Phys. Rev. {\bf D75}, 023523 (2007), arXiv:gr-qc/0612070.

\bibitem{kp-lambda}
W.~Kaminski and T.~Pawlowski,
\newblock Phys. Rev. {\bf D81}, 024014 (2010), arXiv:0912.0162.

\bibitem{ap-lambda}
T.~Pawlowski and A.~Ashtekar,
\newblock Phys. Rev. {\bf D85}, 064001 (2012), arXiv:1112.0360.

\bibitem{madrid-comp}
G.~A. Mena~Marugan, J.~Olmedo, and T.~Pawlowski,
\newblock Phys. Rev. {\bf D84}, 064012 (2011), arXiv:1108.0829.

\bibitem{rad}
T.~Pawlowski, R.~Pierini, and E.~Wilson-Ewing,
\newblock Phys. Rev. {\bf D90}, 123538 (2014), arXiv:1404.4036.

\bibitem{numlsu-2}
P.~Diener, B.~Gupt, and P.~Singh,
\newblock Class. Quant. Grav. {\bf 31}, 105015 (2014), arXiv:1402.6613.

\bibitem{numlsu-3}
P.~Diener, B.~Gupt, M.~Megevand, and P.~Singh,
\newblock Class. Quant. Grav. {\bf 31}, 165006 (2014), arXiv:1406.1486.

\bibitem{recall}
A.~Corichi and P.~Singh,
\newblock Phys. Rev. Lett. {\bf 100}, 161302 (2008), arXiv:0710.4543.

\bibitem{kp-fluc}
W.~Kaminski and T.~Pawlowski,
\newblock Phys. Rev. {\bf D81}, 084027 (2010), arXiv:1001.2663.

\bibitem{cm-fluc1}
A.~Corichi and E.~Montoya,
\newblock Int. J. Mod. Phys. {\bf D21}, 1250076 (2012), arXiv:1105.2804.

\bibitem{cm-fluc2}
A.~Corichi and E.~Montoya,
\newblock Phys. Rev. {\bf D84}, 044021 (2011), arXiv:1105.5081.

\bibitem{ps16a}
P.~Singh,
\newblock Int. J. Mod. Phys. {\bf D25}, 1642001 (2016), arXiv:1604.03828.

\bibitem{we-bianchi9}
E.~Wilson-Ewing,
\newblock Phys. Rev. {\bf D82}, 043508 (2010), arXiv:1005.5565.

\bibitem{pswe}
P.~Singh and E.~Wilson-Ewing,
\newblock Class. Quant. Grav. {\bf 31}, 035010 (2014), arXiv:1310.6728.

\bibitem{ck-b9}
A.~Corichi and A.~Karami,
\newblock Int. J. Mod. Phys. {\bf D25}, 1642011 (2016), arXiv:1605.01383.

\bibitem{bv}
C.~G. Boehmer and K.~Vandersloot,
\newblock Phys. Rev. {\bf D76}, 104030 (2007), arXiv:0709.2129.

\bibitem{cs-schw}
A.~Corichi and P.~Singh,
\newblock Class. Quant. Grav. {\bf 33}, 055006 (2016), arXiv:1506.08015.

\bibitem{bgps-spatial}
B.~Gupt and P.~Singh,
\newblock Phys. Rev. {\bf D85}, 044011 (2012), arXiv:1109.6636.

\bibitem{ks-constant}
N.~Dadhich, A.~Joe, and P.~Singh,
\newblock Class. Quant. Grav. {\bf 32}, 185006 (2015), arXiv:1505.05727.

\bibitem{ks-bound}
A.~Joe and P.~Singh,
\newblock Class. Quant. Grav. {\bf 32}, 015009 (2015), arXiv:1407.2428.

\bibitem{ks-strong}
S.~Saini and P.~Singh,
\newblock (2016), arXiv:1606.04932.

\bibitem{awe-bianchi2}
A.~Ashtekar and E.~Wilson-Ewing,
\newblock Phys. Rev. {\bf D80}, 123532 (2009), arXiv:0910.1278.

\bibitem{ck-closed1}
A.~Corichi and A.~Karami,
\newblock Phys. Rev. {\bf D84}, 044003 (2011), arXiv:1105.3724.

\bibitem{psvt}
P.~Singh and F.~Vidotto,
\newblock Phys. Rev. {\bf D83}, 064027 (2011), arXiv:1012.1307.

\bibitem{ck-closed2}
A.~Corichi and A.~Karami,
\newblock Class. Quant. Grav. {\bf 31}, 035008 (2014), arXiv:1307.7189.

\bibitem{ssd}
P.~Singh, M.~Sami, and N.~Dadhich,
\newblock Phys. Rev. {\bf D68}, 023522 (2003), arXiv:hep-th/0305110.

\bibitem{sst}
M.~Sami, P.~Singh, and S.~Tsujikawa,
\newblock Phys. Rev. {\bf D74}, 043514 (2006), arXiv:gr-qc/0605113.

\bibitem{lqc-bigrip3}
D.~Samart and B.~Gumjudpai,
\newblock Phys. Rev. {\bf D76}, 043514 (2007), arXiv:0704.3414.

\bibitem{lqc-bigrip4}
T.~Naskar and J.~Ward,
\newblock Phys. Rev. {\bf D76}, 063514 (2007), arXiv:0704.3606.

\bibitem{vt}
V.~Taveras,
\newblock Phys. Rev. {\bf D78}, 064072 (2008), arXiv:0807.3325.

\bibitem{ss-inverse1}
P.~Singh and S.~K. Soni,
\newblock Class. Quant. Grav. {\bf 33}, 125001 (2016), arXiv:1512.07473.

\bibitem{st}
P.~Singh and A.~Toporensky,
\newblock Phys. Rev. {\bf D69}, 104008 (2004), arXiv:gr-qc/0312110.

\bibitem{thermal}
J.~Magueijo and P.~Singh,
\newblock Phys. Rev. {\bf D76}, 023510 (2007), arXiv:astro-ph/0703566.

\bibitem{entropy} K.~A.~Meissner,
  Class.\ Quant.\ Grav.\  {\bf 21}, 5245 (2004), arXiv:gr-qc/0407052.

\bibitem{ps6}
P.~Singh,
\newblock Phys. Rev. {\bf D73}, 063508 (2006), arXiv:gr-qc/0603043.

\bibitem{ps05}
P.~Singh,
\newblock Class. Quant. Grav. {\bf 22}, 4203 (2005), arXiv:gr-qc/0502086.

\end{thebibliography}

\end{document}